\documentclass[aps,prd,preprint,groupedaddress,eqsecnum,nofootinbib]{revtex4}
\usepackage{latexsym}
\usepackage{graphicx}
\usepackage{amsmath,amsfonts,amssymb,amsthm,amstext,amscd,eucal}
\usepackage{array}
\usepackage{hyperref}
\usepackage{natbib}

\begin{document}
\title{
\normalsize \mbox{ }\hspace{\fill}\begin{minipage}{12 cm} {\tt
~~~~~~~~~~~~~~~~~~ UPR-1155-T, hep-th/0606001}{\hfill}
\end{minipage}\\[5ex]
{\large\bf Proton decay via dimension-six operators in \\
intersecting D6-brane models
\\[1ex]}}
\date{\today}
\author{Mirjam Cveti{\v c} and Robert Richter}
\affiliation{ Department of Physics and Astronomy, University of Pennsylvania, \\
Philadelphia, PA 19104, USA} \email{cvetic@cvetic.hep.upenn.edu;
rrichter@physics.upenn.edu}
\date{\today}

\begin{abstract}
We analyze the proton decay via dimension six operators in
supersymmetric $SU(5)$-Grand Unified models based on intersecting
D6-brane constructions in Type IIA string theory orientifolds. We
include in addition to $\mathbf{10^{^{*}}} \mathbf{10}
\mathbf{10}^{^{*}} \mathbf{10}$ interactions also the operators
arising from $\mathbf{\bar{5}^{^{*}}} \mathbf{\bar{5}}
\mathbf{10}^{^{*}} \mathbf{10}$ interactions.  We provide  a
detailed construction of  vertex operators for  any massless  string
excitation arising for  arbitrary intersecting D-brane
configurations in Type IIA toroidal orientifolds. In  particular, we
provide explicit string vertex operators for the $\mathbf{10}$ and
$\mathbf{\bar{5}}$ chiral superfields  and calculate explicitly the
string theory correlation functions for above operators. In the
analysis we chose the most symmetric configurations in order to
maximize proton decay rates for the above dimension six operators
and  we obtain a small enhancement relative to  the field theory
result.
After relating the string proton decay rate to field theory
computations the string contribution to the proton lifetime is
$\tau^{ST}_p =(0.5-2.1)\times 10^{36} years$, which could be up to a
factor of three shorter than that predicted in field theory.\

\end{abstract}
\maketitle

\tableofcontents
\section{Introduction}
Grand unified theories (GUT's) \cite{Georgi:1974sy} not only give a
neat and aesthetic description of our four dimensional world but
also lead to an explanation of electric charge quantization and -
with the aid of supersymmetry - predict the value of
$\sin^2\theta_W$ in very good agreement with the experimental one.
Moreover GUT's lead to Baryon number violating processes; in
particular they predict proton decay \cite{Langacker:1980js} (for
a recent review on proton decay see \cite{Nath:2006ut}). \\
In supersymmetric GUT field theories
\cite{{Sakai:1981gr},{Dimopoulos:1981zb}} the proton decay can occur
either by an exchange of a super heavy SUSY particle which
corresponds to a decay via the dimension $5$ operator $\int
d^2\theta\, Q^3\,L$ or by a super heavy gauge boson
exchange\footnote{We forbid proton decay due to dimension four
operators by introducing $R$-symmetry.}. The latter corresponds to a
decay via the dimension $6$ operator $\int d^4\theta\,
Q^2\,\tilde{Q}^{*}\,\tilde{L}^{*}$. In the simplest supersymmetric
GUT models, proton decay mediated via dimension $5$ operators
dominates and recent computations predict a lifetime for the proton,
which is below the present experimental bounds
\cite{{Murayama:2001ur},{Dermisek:2000hr},{Hisano:2000dg}}, but
\cite{{Emmanuel-Costa:2003pu},{Bajc:2002pg}}. The fact that proton
decay has not yet been observed, suggests the existence of some
mechanism that suppresses or even forbids these dimension $5$
operators \cite{{Bajc:2002bv},{Bajc:2002pg}}, so that after all the
proton decay via dimension six
operators \cite{FileviezPerez:2004hn} is the most dominant one. \\
In this paper we investigate proton decay via dimension six
operators in supersymmetric GUT models based on intersecting
D6-brane constructions on type IIA string theory orientifolds. More
precisely, we compute the string effects on the proton's decay into
a pion and a positron ($p\rightarrow \pi^0 e^+$) for supersymmetric
$SU(5)$-GUT-like models arising from intersecting D6-brane
constructions. In $SU(5)$-GUT's there are two different amplitudes
that contribute to this proton decay rate:  $\langle
\mathbf{10^{^{*}}} \mathbf{10} \mathbf{10}^{^{*}} \mathbf{10}
\rangle $  and $\langle \mathbf{\bar{5}^{^{*}}} \mathbf{\bar{5}}
\mathbf{10}^{^{*}} \mathbf{10}\rangle $, where $\mathbf{\bar{5}}$
and $\mathbf{10}$ denote the multiplets of the gauge group $SU(5)$.
For intersecting D6-brane  constructions with supersymmetric
$SU(5)$-GUT's
\cite{{Cvetic:2001nr},{Cvetic:2001tj},{Cvetic:2002pj}}\footnote{For
a review on intersecting D-brane constructions see,
 e.g.,
\cite{Blumenhagen:2005mu}; for the original work on
non-supersymmetric intersecting D-branes, see
\cite{{Blumenhagen:2000wh},{Aldazabal:2000cn}, {Blumenhagen:2001te},
{Aldazabal:2000dg}}, and chiral supersymmetric ones ,
see\cite{{Cvetic:2001tj}, {Cvetic:2001nr}} and also
\cite{Angelantonj:2000hi}. For flipped $SU(5)$ constructions see
\cite{{Chen:2005ab}, {Chen:2005cf}}. For supersymmetric $SU(5)$ GUT
constructions within Type II rational conformal field theories see:
\cite{{Dijkstra:2004cc},{Anastasopoulos:2006da}} and references
therein. For a related study on Calabi Yau manifolds see
\cite{Tatar:2006dc}.}, the latter amplitude was computed in
\cite{Klebanov:2003my}, by explicitly calculating the string
amplitude contribution  to $ \mathbf{10^{^{*}}} \mathbf{10}
\mathbf{10}^{^{*}} \mathbf{10}$ operators. However, even after
pushing all the parameters to the limit, in order to maximize the
proton decay rate, the string contribution to it is at most
comparable to the field theory one.
In this work, we explicitly evaluate the amplitude $\langle
\mathbf{\bar{5}^{^{*}}} \mathbf{\bar{5}} \mathbf{10}^{^{*}}
\mathbf{10}\rangle$ in the same class of models.
\\
As in \cite{Klebanov:2003my}, instead of performing the calculation
in a specific model, we rather use generic universal features of
intersecting D-brane model constructions which are relevant for
determining the proton decay rate. In general, the amplitude is
sensitive to the local structure of the intersection and the way the
D6-branes are wrapped around the compact space. Assuming that the
size of the compactified volume is bigger than the string size, the
latter effects can be neglected and the computation can be performed
for  a local D6-brane configuration where we do not need to worry
about the embedding in the compact space. This approach allows us to
make predictions about the proton decay rate in a general class of
intersecting D6-brane orientifold models. In generic models the
matter fields $\mathbf{10}$ and $\mathbf{\bar{5}}$ are not located
at the same intersection, which leads to an overall suppression of
the amplitude $\langle \mathbf{\bar{5}^{^{*}}} \mathbf{\bar{5}}
\mathbf{10}^{^{*}} \mathbf{10}\rangle$. In this work, in order to
maximize the effect, we assume the most symmetric case that all the
matter arises at intersections that are on top of each other.
Therefore, we rather compute an upper bound for the string
contribution to the proton decay rate in these models than
determining the complete amplitude which is model dependent.\\
This paper is organized as follows. In section 2 we describe the
local setup in which we work and derive the conditions on the
intersecting angles, in order to obtain the matter fields in the
representation $\mathbf{10}$ and $\mathbf{\bar{5}}$ at the
intersection, simultaneously. In section 3 we apply the
prescription, given in appendix A, to construct the vertex operators
for the matter fields. Section 4 is dedicated to the computation of
the string scattering amplitudes, including their normalization.
Section 5 states the results of the numerical analysis, while the
details can be found in appendix B. In section 6 we relate the
string theory results to the four-dimensional field theory and
determine the implication of the string scattering amplitude to the
proton lifetime. Finally in section 7 we present our conclusions. In
appendix A we give a detailed description, how to construct properly
vertex operators for strings stretched between two intersecting
D-branes.
\section{Setup}
We want to analyze proton decay which occurs due to dimension $6$
operators in a local intersecting D6-brane configuration. Therefore,
we have to consider scattering amplitudes of the form $\langle
\mathbf{\bar{5}^{^{*}}} \mathbf{\bar{5}} \mathbf{10}^{^{*}}
\mathbf{10}\rangle $ and $\langle \mathbf{10^{^{*}}} \mathbf{10}
\mathbf{10}^{^{*}} \mathbf{10} \rangle $, where $\mathbf{\bar{5}}$
and $\mathbf{10}$ denote the multiplets of the gauge group $SU(5)$.
While the latter amplitude was already examined in
\cite{Klebanov:2003my}, we will determine the additional
contribution to the proton decay arising from the amplitude $\langle
\mathbf{\bar{5}^{^{*}}}
\mathbf{\bar{5}} \mathbf{10}^{^{*}} \mathbf{10}\rangle $. \\
Since we shall only consider scattering arising at the local
intersection, the first step is to derive conditions on the angles
so that we have at the local intersection matter fields in the
$\mathbf{\bar{5}}$ and $\mathbf{10}$ representation, simultaneously.
We will show that this condition is satisfied only for particular
regions. For the explicit analysis we shall employ the toroidal
orientifold construction and take the size of the tori larger than
the inverse string tension, thus suppressing effects due to the
world-sheet instantons.  In this limit we shall calculate the
four-point string amplitudes for the chiral superfields at the
D6-brane intersections at the origin of the toroidal orientifold. In
that sense the analysis can be applied as the leading order
calculation of string amplitudes for the states at the same D6-brane
intersection within any orientifold construction.\\
Let us briefly review the main properties of intersecting D6-models.
Generically, one has a number of stacks of D6-branes ($N_i$ denotes
the number of D-branes for the $i$-th stack), which fill the
four-dimensional Minkowski space and intersect each other in the
internal space. Open string excitations located at the intersections
correspond to four-dimensional chiral fermions transforming in the
bifundamental representation $(N_i,\bar{N}_j)$, while open strings
starting and ending at the same stack of D6-branes transform as
seven-dimensional $U(N_i)$ gauge bosons. In order to make contact
with the real world, one has to compactify the six-dimensional
internal space which leads to additional consistency conditions on
the model called the RR tadpole conditions. D-branes act as sources
for the Ramond-Ramond (RR)-charges which need to be canceled due to
Gauss' law in the internal compact space
\cite{{Blumenhagen:2000wh},{Gimon:1996rq}}. Typically one introduces
Orientifold six (O6-) planes, not only because they carry negative
RR-charge, but also because they can maintain supersymmetry in the
four-dimensional world, while the introduction of anti-D-branes
would break all the supersymmetry. The orientifold action leads to
image $\text{D6}^\prime$-branes and open strings stretched between a
D6-brane and its image transform as symmetric or anti-symmetric
representation of $U(N_i)$. As mentioned in the introduction, we
rather investigate the proton decay amplitude in a local D6-brane
configuration than in a specific model. In the following we discuss
all the necessary ingredients for this configuration to obtain a
supersymmetric $SU(5)$-GUT like model \cite{Cvetic:2002pj} (for the
non-supersymmetric case see
\cite{{Blumenhagen:2001te}, {Axenides:2003hs}}).\\
As explained above the analysis of the D-brane configuration we
compactify the internal dimensions are on a factorizable six-torus
$T^6$. Later we assume that the compactification volume is larger
than the string scale so that local effects dominate the amplitude
and global ones can be neglected. This assumption also allows us to
embed the local D-brane configuration, described below, into an
arbitrary compactification manifold. \\
The complex coordinates of the factorizable six-torus $T^6=T^2
\times T^2\times T^2 $ are given by
\begin{align*}
z_1=x^4+\mathrm{i}x^5 \qquad z_2=x^6+\mathrm{i}x^7 \qquad
z_1=x^8+\mathrm{i}x^9.
\end{align*}
In order to construct an $SU(5)$ GUT model we shall consider very symmetric
configurations of D6-branes. We take a stack $b$ of $M$ D6-branes oriented in
the 0123468 directions that coincides  with a stack $a$ of $5$ D6 branes along
the 0123 directions and forms (supersymmetric) intersecting angles  with stack
$b$ in the internal toroidal directions. The dimensions 0123 have an
interpretation as a $3+1$ dimensional intersecting brane world. Both types of
D-branes are wrapped on the $(n^I,m^I)$ cycle of the $I^{th}$ torus.
Obviously, the wrapping numbers of the stack $b$ are given by
\begin{align}
b:\,\,\,(n^1_b,m^1_b) (n^2_b,m^2_b) (n^3_b,m^3_b)=(1,0) (1,0) (1,0),
\label{intersection number b}
\end{align}
while the one from stack $a$ can take the general form
\begin{align}
a:\,\,\,(n^1_a,m^1_a) (n^2_a,m^2_a) (n^3_a,m^3_a).
\label{intersection number a}
\end{align}
Given the wrapping numbers, one can compute the intersection angles
which  are in general given by ($R^I_1$,$R^I_2$ denote the radii of
the $I^{th}$ torus)\footnote{Note that with this definition
clockwise angles are positive and counter-clockwise negative.}
\begin{align*}
\theta^{I}_{ab}=\theta^{I}_{a}-\theta^{I}_{b}=
\arctan{\left(\frac{m^I_a R^I_2}{n^I_a
R^I_1}\right)}-\arctan{\left(\frac{m^I_b R^I_2}{n^I_b R^I_1}\right)}
\end{align*}
and in our case take the simple form (since $\theta_b=0$)
\begin{align}
\theta^{I}_{ab}= \arctan{\left(\frac{m^I_a R^I_2}{n^I_a
R^I_1}\right)}. \label{angle theta}
\end{align}
In order to cancel the RR-tadpoles, we must introduce O6-planes and
in particular the orientifold action $\Omega R$, where $\Omega$ is
the world-sheet parity and $R$ acts by
\begin{align*}
R : (z_1,z_2,z_3) \rightarrow (\bar{z}_1,\bar{z}_2,\bar{z}_3).
\end{align*}
This orientifold action forces us to include stacks of image
D-branes. Since we chose stack $b$ to lie on top of the orientifold
$O6$-plane, it is invariant under  the orientifold action: for $M$
coincident branes on top of the $O6$-plane the $\Omega R$ projection
leads to the gauge group $Sp(2M)$. For the stack $a$ we have to
introduce an image stack $a'$ of $5$ D6-branes whose wrapping
numbers are given by
\begin{align}
a':\,\,\,(n^1_a,-m^1_a) (n^2_a,-m^2_a) (n^3_a,-m^3_a)\,\, .
\label{intersection number a'}
\end{align}
Fermions that arise from strings stretched between $a$ and $a'$
transform in the antisymmetric representation of $SU(5)\times
SU(5)$, due to the fact that the D-branes intersect at the origin of
the torus. Depending on the sign of the intersection number these
fermions transform as $\mathbf{10}$'s or $\mathbf{\overline{10}}$'s
. Fermions in the $ab$ and $ab'$ sector transform in the
bifundamental representation $(\mathbf{5},\mathbf{M})$ or
$(\mathbf{\bar{5}},\mathbf{M})$\footnote{$\mathbf{M}$ denotes the
representation of the gauge group $Sp(2M).$} again depending on the
sign of the intersection number. In general, the intersection number
for two intersecting D-branes $a$ and $b$ is given by
\begin{align}
I_{ab}=\prod^3_{I=1} \, \left(n^I_a m^I_b - m^I_a n^I_b\right)\,\,.
\label{general Intersection number}
\end{align}
Now we have all the ingredients to determine the conditions the
intersection angles $\theta_I$ have to satisfy in order to observe
matter fields transforming as $\mathbf{\bar{5}}$ and $\mathbf{10}$
at the intersection, simultaneously. Using \eqref{intersection
number b}, \eqref{intersection number a}, \eqref{intersection number
a'} and \eqref{general Intersection number} we obtain for the
intersection numbers $I_{ab}$ and $I_{aa'}$
\begin{align}
I_{ab}=(-1)^3 \, \prod^3_{I=1} \,  m^I_a \qquad I_{aa'}=(-2)^3\,
\prod^3_{I=1} \,  n^I_a m^I_a\,\,. \label{intersection number model}
\end{align}
Obviously, the sign of the intersection number depends on the sign
of the wrapping numbers. For every angle $\theta_I$ (from now on we
denote $\theta^I_{ab}$ by $\theta_I$ where $\theta^I_{ab}$ is given
by \eqref{angle theta}) we have to distinguish between four
different cases
\begin{align}
\label{angles wrapping number}
 \nonumber &\bullet \,\,n^I_a, \,m^I_a > 0
\,\,\, \textrm{which corresponds to an angle with} \,\,\,
0<\theta_I<\frac{\pi}{2} \\
\nonumber &\bullet \,\,n^I_a > 0,\,\, m^I_a < 0 \,\,\, \textrm{which
corresponds
to an angle with} \,\,\,-\frac{\pi}{2}<\theta_I<0 \qquad \\
&\bullet \,\,n^I_a<0  \,\,\, m^I_a > 0 \,\,\, \textrm{which
corresponds to an angle with} \,\,\, \frac{\pi}{2}<\theta_I<\pi \\
\nonumber &\bullet \,\,n^I_a, \, m^I_a < 0 \,\,\, \textrm{which
corresponds to an angle with\,\,\,
$-\frac{\pi}{2}<\theta_I<-\pi$}\,\,.
\end{align}
Since we want to analyze proton decay in a supersymmetric GUT model
the choice of the intersection angles $\theta_I$ is not arbitrary;
the sum has to satisfy \cite{Berkooz:1996km}
\begin{align}
\theta_1+\theta_2+\theta_3=0 \qquad \mod 2\pi \,\,.\label{SUSY
condition}
\end{align}
This requirement restricts the choice of the angles. First we
consider the case that the angles add up to $0$ and later on we also
analyze the configuration where the sums of the angles are $2\pi$ or
$-2\pi$. If the sum is equal to $0$ then one or two of the angles
have to be negative. If only one angle is negative, let us assume
without loss of generality that $\theta_3<0$. Since for all angles
$|\theta_I| \leq \pi$, we distinguish between four different cases
for which we obtain, by applying \eqref{intersection number model}
and \eqref{angles wrapping number}, the intersection numbers and in
particular their signs
\begin{itemize} \item{ $I_{ab}>0$ \,and\, $I_{aa'}>0$ \,for
\,\,  $ 0<\theta_1<\frac{\pi}{2}\qquad
 0<\theta_2<\frac{\pi}{2} \qquad  -\frac{\pi}{2} <\theta_3< 0$}
\item{$I_{ab}>0$ \,and\, $I_{aa'}<0$ \,for \,\,
$0<\theta_1<\frac{\pi}{2} \qquad
 0<\theta_2<\frac{\pi}{2} \qquad -\pi <\theta_3< -\frac{\pi}{2}$}
\item{$I_{ab}>0$ \,and \,$I_{aa'}>0$ \,for \,\, $\frac{\pi}{2}<
\theta_1<\pi \qquad
 0<\theta_2<\frac{\pi}{2} \qquad -\pi <\theta_3< -\frac{\pi}{2}$}
\item{$I_{ab}>0$ \,and\, $I_{aa'}>0$ \,for \,\,
$0<\theta_1<\frac{\pi}{2} \qquad \frac{\pi}{2} <\theta_2<\pi \qquad
-\pi <\theta_3< -\frac{\pi}{2}$}\,\,.
\end{itemize}
For all combinations of $\theta_I$'s that fulfill the above stated
properties ($\sum \theta_I =0$ and one angle is negative) we see
that the strings stretched between D-branes $a$ and $b$ transform as
$\mathbf{5}$ instead of the desired $\mathbf{\bar{5}}$. Therefore we
do not observe a 4-point interaction of the form
$\mathbf{\bar{5}}^{*}\mathbf{\bar{5}}
\mathbf{10}^{*} \mathbf{10}$ at the intersection. \\
Analyzing the case of two negative angles (without loss of
generality we assume that $\theta_1$ and $\theta_2$ are negative) we
again distinguish between four different cases
\begin{itemize}
\item{ $I_{ab}<0$ \,and\, $I_{aa'}<0$ \,for \,\,  $
-\frac{\pi}{2}<\theta_1<0 \qquad -\frac{\pi}{2} <\theta_2< 0 \qquad
0<\theta_1<\frac{\pi}{2}$} \item{$I_{ab}<0 $ \,and\, $I_{aa'}>0$
\,for \,\, $ -\frac{\pi}{2}<\theta_1<0 \qquad -\frac{\pi}{2}
<\theta_2< 0 \qquad \frac{\pi}{2}< \theta_3<\pi$}
\item{$I_{ab}<0$ \,and \,$I_{aa'}<0$ \,for \,\, $
 -\pi<\theta_1<-\frac{\pi}{2} \,\,\,\,\,\,\, -\frac{\pi}{2} <\theta_2< 0 \qquad \frac{\pi}{2}<
\theta_3<\pi $} \item{$I_{ab}<0$ \,and\, $I_{aa'}<0$ \,for \,\, $
-\frac{\pi}{2}<\theta_1<0 \qquad -\pi<\theta_2<-\frac{\pi}{2} \qquad
\frac{\pi}{2}< \theta_3<\pi$}\,\,.
\end{itemize}
Only in the region
$-\frac{\pi}{2}<\theta_{1,2}<0\,,\frac{\pi}{2}<\theta_{3}<\pi$ we
observe matter fields transforming as $\mathbf{\bar{5}}$ and
$\mathbf{10}$, where strings stretched between the D-branes $a$ and
$b$ transform as $\mathbf{\bar{5}}$ and strings
stretched between $a$ and $a'$ transform as $\mathbf{10}$.\\
Let us now turn to the case in which the intersection angles
$\theta_I$ add up to $2\pi$. Then all the angles are positive and we
have to distinguish between three different configurations (without
loss of generality let us assume that $\theta_1$ is always bigger
than $\frac{\pi}{2}$)
\begin{itemize}
\item{$I_{ab}<0$ \,and\, $I_{aa'}<0$ \,for \,\, $\frac{\pi}{2}<
\theta_1<\pi \qquad \frac{\pi}{2}<\theta_2<\pi \qquad 0 <\theta_3<
\frac{\pi}{2} $} \item{$I_{ab}<0$ \,and \,$I_{aa'}<0$ \,for \,\,
$\frac{\pi}{2}< \theta_1<\pi \qquad
 0<\theta_2<\frac{\pi}{2}\qquad \frac{\pi}{2} <\theta_3< \pi $}
 \item{$I_{ab}<0$ \,and \,$I_{aa'}>0$ \,for \,\, $\frac{\pi}{2}< \theta_1<\pi \qquad
\frac{\pi}{2}<\theta_2<\pi\qquad \frac{\pi}{2} <\theta_3< \pi
$}\,\,.
\end{itemize}
Again only in one region, $\frac{\pi}{2}<\theta_{1,2,3}<\pi$, we
observe matter fields transforming as $\mathbf{\bar{5}}$ and
$\mathbf{10}$, where strings stretched between the D-branes $a$ and
$b$ transform as $\mathbf{\bar{5}}$ and strings stretched
between $a$ and $a'$ transform as $\mathbf{10}$. \\
Finally, we examine the case in which the angles add up to $-2\pi$.
Here all three angles have to be negative and again one has to
distinguish between three different cases (without loss of
generality we assume that $\theta_1$ is smaller than
$-\frac{\pi}{2}$)
\begin{itemize}
\item{$I_{ab}>0$ \,and\, $I_{aa'}>0$ \,for \,\,
$-\pi<\theta_1<\frac{\pi}{2} \qquad -\pi<\theta_2<\frac{\pi}{2}
\qquad -\frac{\pi}{2} <\theta_3< 0 $} \item{$I_{ab}>0$
\,and\,\,\,$I_{aa'}>0$ \,for \,\, $-\pi<\theta_1<\frac{\pi}{2}
\qquad
 -\frac{\pi}{2}<\theta_2<0 \qquad -\pi<\theta_3< -\frac{\pi}{2} $}
 \item{$I_{ab}>0$ \,and \,\,$I_{aa'}<0$ \,for \,\, $-\pi<
 \theta_1<-\frac{\pi}{2}\,\,\,\,\,\,\,
-\pi<\theta_2<-\frac{\pi}{2}\qquad -\pi <\theta_3< -\frac{\pi}{2}
$}\,\,.
\end{itemize}
As in the first case, the analysis shows that strings stretched
between D-branes $a$ and $b$ transform as $\mathbf{5}$ under the
$U(5)$ gauge group. Therefore, at the intersection we do not have
any matter fields transforming as $\mathbf{\bar{5}}$.\\
Summarizing, we determined that only for the two regions
$-\frac{\pi}{2}<\theta_{1,2}<0\,,\frac{\pi}{2}<\theta_{3}<\pi$ and
$\frac{\pi}{2}<\theta_{1,2,3}<\pi$ we have matter fields
transforming as $\mathbf{\bar{5}}$ and $\mathbf{10}$ at the
intersection simultaneously. In addition to the amplitude $\langle
\mathbf{10^{^{*}}} \mathbf{10} \mathbf{10}^{^{*}} \mathbf{10}
\rangle $, we have for these two regions only, a non-suppressed
contribution from $\langle \mathbf{\bar{5}^{^{*}}} \mathbf{\bar{5}}
\mathbf{10}^{^{*}} \mathbf{10}\rangle $ to the proton decay rate. In
order to compute these two amplitudes we need the corresponding
vertex operators to the states $\mathbf{\bar{5}}$, and $\mathbf{10}$
in the respective configurations, which we determine in the next
section.
\section{Vertex Operators}

For different D-brane configurations we have different vacua and
therefore different vertex operators. Knowing the D-brane
configuration we can use the prescription given in appendix A to
obtain the vertex operator for the massless fermion in the R-sector.
In this way we can easily determine the vertex operators for
$\mathbf{\bar{5}}$, arising from strings stretched between the
stacks $a$ and $b$. The vertex operator for $\mathbf{10}$ requires
more effort. The simple approach just to replace the $\theta_I$ in
the $\mathbf{\bar{5}}$ vertex operator by the double, $2\theta_I$
only works for $|\theta_I|<\frac{1}{2}$\footnote{From now on we
replace $\theta_I$ by $\theta_I/\pi$ so that $\theta_I \, \epsilon
\, [-1,1].$}, since in the expansion of the bosonic
\eqref{modeexpansionintersecting bosonic} and fermionic degrees of
freedom \eqref{modeexpansionintersecting fermionic} the shift number
$\theta_I$ has to be in the interval $[-1,1]$. Therefore if
$\theta_I>\frac{1}{2}$ we need to find an expression $\nu_I$ which
lies between $0$ and $1$ and describes the D-brane configuration
$aa'$.
\begin{figure}
  \includegraphics[width=140mm]{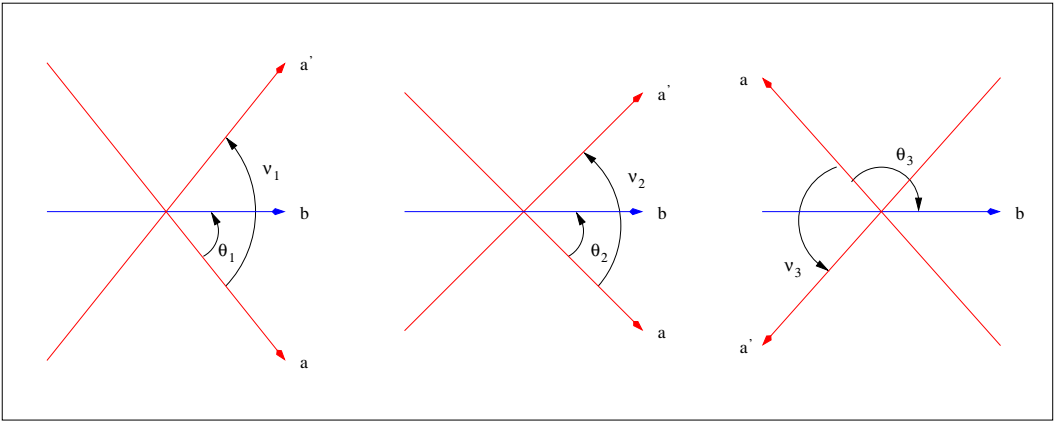}\\
  \caption{Intersection angles for the case $-\frac{1}{2}<\theta_{1}<0 ,
  -\frac{1}{2}<\theta_{2}<0, \frac{1}{2}<\theta_{3}<1$. }\label{figure 1}
\end{figure}
Figure \ref{figure 1} which shows the D-brane configuration for the
case $-\frac{1}{2}<\theta_{1,2}<0\,, \frac{1}{2}<\theta_3<1$. The
vertex operator in the $(-\frac{1}{2})$-ghost picture for the
massless fermion, arising from a string stretched between D-branes
$a$ and $b$ is given by (keep in mind that $\theta_{1,2}$ are
negative)
\begin{align}
V^{-\frac{1}{2}}_{\bar{5}}(z) &= \Lambda^{\bar{5}} \,
\mathrm{e}^{-\frac{\phi}{2}(z)}\,S^{\alpha}(z) \prod^2_{I=1}
\sigma_{-\theta_I} (z)\,
\mathrm{e}^{-\mathrm{i}\left(\theta_I+\frac{1}{2}\right) H_I(z) } \,
\sigma_{1-\theta_3} (z)\,
\mathrm{e}^{-\mathrm{i}\left(\theta_3-\frac{1}{2}\right) H_3(z) } \,
\mathrm{e}^{\mathrm{i} k \cdot X (z)}.
\end{align}
Now we turn to the $aa'$ sector in which the string state transforms
as $\mathbf{10}$. We see that the intersection angle in the third
complex dimension is given by $\nu_{3}=-2+2\theta_{3}$. Note that
the intersection angle $\nu_{3}$ is negative and lies between $-1$
and $0$, since $\theta_{3}$ takes a value between $\frac{1}{2}$ and
$1$ and therefore the corresponding vertex operator for the state
$\mathbf{10}$ takes the form
\begin{align}
V^{-\frac{1}{2}}_{10}(z) & = \Lambda^{10} \,
\mathrm{e}^{-\frac{\phi}{2}(z)}\,S^{\alpha}(z) \prod^3_{I=1}
\sigma_{1+\nu_I} (z)\,
\mathrm{e}^{\mathrm{i}\left(\nu_I+\frac{1}{2}\right) H_I(z) } \,
\mathrm{e}^{\mathrm{i} k \cdot X (z)},
\end{align}
where the angles $\nu_I$ are given by
\begin{align*}
\nu_1=2\theta_1 \qquad \nu_2=2\theta_2 \qquad
\nu_3=-2+2\theta_3\,\,.
\end{align*}
Notice, that the angles $\nu_I$ add up to $-2$ so that the SUSY
condition \eqref{SUSY condition} is satisfied. In an analogous way
(look at figure \ref{figure 2}), we obtain for the other D-brane
configuration
\begin{figure}
  \includegraphics[width=140mm]{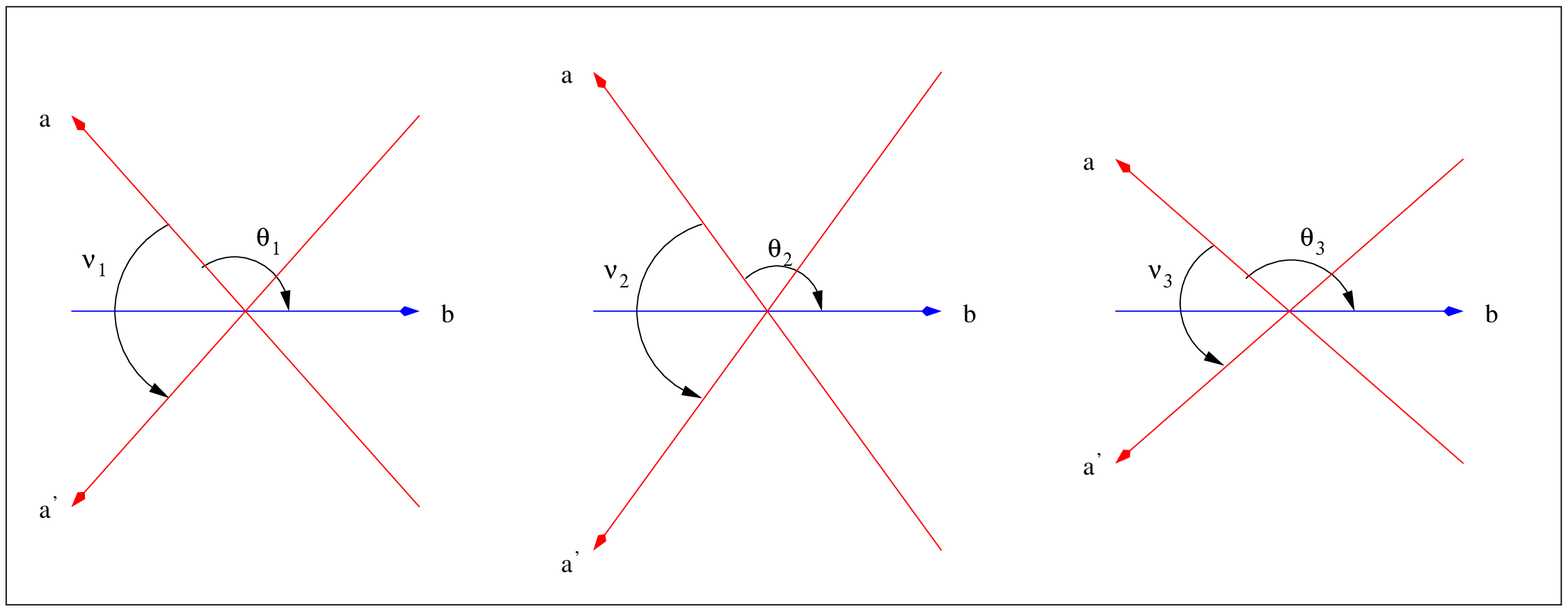}\\
  \caption{Intersection angles for the case $\frac{1}{2}<\theta_{1}<1,
  \frac{1}{2}<\theta_{2}<1,
\frac{1}{2}<\theta_{3}<1$ }\label{figure 2}
\end{figure}

\begin{itemize}
\item{$\frac{1}{2}<\theta_{1}<1\qquad \frac{1}{2}<\theta_{2}<1
\qquad \frac{1}{2}<\theta_{3}<1$\\
\\
For this configuration the vertex operator that creates a string
stretched between $a$ and $b$ is
\begin{align}
V^{-\frac{1}{2}}_{\bar{5}}(z) & = \Lambda^{\bar{5}} \,
\mathrm{e}^{-\frac{\phi}{2}(z)}\,S^{\alpha}(z) \prod^3_{I=1}
\sigma_{1-\theta_I} (z)\,
\mathrm{e}^{-\mathrm{i}\left(\theta_I-\frac{1}{2}\right) H_I(z) } \,
\mathrm{e}^{\mathrm{i} k \cdot X (z)}\,\,.
\end{align}
The intersection angles $\nu_I$ are given by
\begin{align*}
\nu_1=-2+2\theta_1 \qquad \nu_2=-2+2\theta_2 \qquad
\nu_3=-2+2\theta_3\,\,.
\end{align*}
Obviously, they are all negative, so that the vertex operator which
describes the massless $aa'$-string in the R-sector takes the form
\begin{align}
V^{-\frac{1}{2}}_{10}(z) & = \Lambda^{10} \,
\mathrm{e}^{-\frac{\phi}{2}(z)}\,S^{\alpha}(z) \prod^3_{I=1}
\sigma_{1+\nu_I} (z)\,
\mathrm{e}^{\mathrm{i}\left(\nu_I+\frac{1}{2}\right) H_I(z) } \,
\mathrm{e}^{\mathrm{i} k \cdot X (z)}\,\,.
\end{align}
Again the angles $\nu_I$ add up to $-2$.
 }
\end{itemize}
In order to calculate scattering amplitudes we also need the vertex
operators for $\mathbf{\bar{5}}^*$ and $\mathbf{10}^*$. We obtain
them by replacing the spin field by the spin field with opposite
chirality and at the same time sending the angles $\theta_I$ and
$\nu_I$ to $1-\theta_I$ and $1-\nu_I$, respectively (for negative
angle we replace $\theta_I$ and $\nu_I$ by $-1-\theta_I$ and
$-1-\nu_I$, respectively). For these two cases we obtain
\begin{itemize}
\item{$-\frac{1}{2}<\theta_{1}<0\qquad -\frac{1}{2}<\theta_{2}<0
\qquad \frac{1}{2}<\theta_{3}<1$\\
\\
\begin{align}
V^{-\frac{1}{2}}_{\bar{5}^*}(z) &= {\Lambda^{\bar{5}}}^{\dagger}\,
\mathrm{e}^{-\frac{\phi}{2}(z)}\,\tilde{S}_{\dot{\alpha}}(z)
\prod^2_{I=1} \sigma_{1+\theta_I} (z)\,
\mathrm{e}^{\mathrm{i}\left(\theta_I+\frac{1}{2}\right) H_I(z) } \,
\sigma_{\theta_3} (z)\,
\mathrm{e}^{\mathrm{i}\left(\theta_3-\frac{1}{2}\right) H_3(z) } \,
\mathrm{e}^{\mathrm{i} k \cdot X (z)}
\end{align}
for $\mathbf{\bar{5}}^*$ and
\begin{align}
V^{-\frac{1}{2}}_{10^{*}}(z) & = {\Lambda^{10}}^{\dagger}\,
\mathrm{e}^{-\frac{\phi}{2}(z)}\,\tilde{S}_{\dot{\alpha}}(z)
\prod^3_{I=1} \sigma_{-\nu_I} (z)\,
\mathrm{e}^{-\mathrm{i}\left(\nu_I+\frac{1}{2}\right) H_I(z) } \,
\mathrm{e}^{\mathrm{i} k \cdot X (z)}
\end{align}
for $\mathbf{10}^*$.} \item{$\frac{1}{2}<\theta_{1}<1\qquad
\frac{1}{2}<\theta_{2}<1 \qquad \frac{1}{2}<\theta_{3}<1$\\
\\
\begin{align}
V^{-\frac{1}{2}}_{\bar{5}^{*}}(z) & =
{\Lambda^{\bar{5}}}^{\dagger}\,
\mathrm{e}^{-\frac{\phi}{2}(z)}\,\tilde{S}_{\dot{\alpha}}(z)
\prod^3_{I=1} \sigma_{\theta_I} (z)\,
\mathrm{e}^{-\mathrm{i}\left(\theta_I-\frac{1}{2}\right) H_I(z) } \,
\mathrm{e}^{\mathrm{i} k \cdot X (z)}
\end{align}
for $\mathbf{\bar{5}}^{*}$ and
\begin{align}
V^{-\frac{1}{2}}_{{10}^{*}}(z) & = {\Lambda^{10}}^{\dagger}\,
\mathrm{e}^{-\frac{\phi}{2}(z)}\,\tilde{S}_{\dot{\alpha}}(z)
\prod^3_{I=1} \sigma_{-\nu_I} (z)\,
\mathrm{e}^{-\mathrm{i}\left(\nu_I+\frac{1}{2}\right) H_I(z) } \,
\mathrm{e}^{\mathrm{i} k \cdot X (z)}
\end{align}
 for $\mathbf{10}^*$ .}
\end{itemize}
Finally, we will discuss the Chan-Paton factors. In a setup without
orientifolds strings transform in the bifundamental of $U(N)\times
U(M)$. As already mentioned above, the introduction of orientifolds
changes the transformation behavior. The full orientifold action on
the Chan-Paton factors takes the form
\begin{align*}
\Lambda= -\gamma_{\Omega R} \Lambda^{T} \gamma^{-1}_{\Omega R}\,\,,
\end{align*}
where $\gamma_{\Omega R}$ is given by \cite{Cvetic:2004nk}
\begin{align}
\gamma_{\Omega R}=\left(
\begin{array}{cccc}
  0 & 1_{N}  & 0 & 0 \\
  1_{N} & 0& 0 & 0 \\
  0 & 0 & 0 & 1_{M} \\
  0 & 0 & 1_{M} & 0 \\
\end{array}
\right)\,\,.
\end{align}
The choice of $N=5$ leads to the following Chan-Paton factors for
the $\mathbf{10}$'s
\begin{align}
\Lambda^{10}=\left(
\begin{array}{cccc}
  0 & \lambda_{10} & 0 & 0 \\
  \lambda^T_{10} & 0& 0 & 0 \\
  0 & 0 & 0 & 0 \\
  0 & 0 & 0 & 0 \\
\end{array}
\right) \,\,,\label{chan paton's for the 10's}
\end{align}
where $\lambda_{10}$ is an antisymmetric $5 \times 5$ matrix. For
$M$ we choose $1$ that leads to a $Sp(2)$ gauge group on the D-brane
$b$ which has two components in the fundamental representation. One
component is associated with the matter field $\mathbf{5}$ while the
other corresponds to the Higgs particle. Their Chan-Paton factors
take the form
\begin{align}
\Lambda^{\bar{5}}=\left(
\begin{array}{cccc}
  0 & 0 & 0 & 0 \\
  0 & 0& \lambda_{\bar{5}} &0 \\
  0 & 0 & 0 & 0 \\
  -\lambda^{T}_{\bar{5}} & 0 & 0 & 0 \\
\end{array}
\right) \qquad \qquad \Lambda_{H}=\left(
\begin{array}{cccc}
  0 & 0 & 0 & 0 \\
  0 & 0& 0 &H \\
  -H^{T} & 0 & 0 & 0 \\
  0 & 0 & 0 & 0 \\
\end{array}
\right). \label{chan paton for the 5's}
\end{align}
Here $\lambda_{\bar{5}}$ and $H$ are a $5\times 1$ matrices.
$\lambda_{10}$ and $\lambda_{\bar{5}}$ denote the usual $10$- and
$5$-dimensional representations of the $SU(5)$ gauge group and $H$
is the $5$ dimensional Higgs field in the gauge field theory.
\section{String Amplitude}
Having derived the vertex operators in the previous section, we have
all the ingredients to compute the scattering amplitudes. Assuming
that the compactification volume is larger than the string scale
worldsheet instantons are suppressed and it is sufficient to compute
just the quantum part of the amplitudes. First we will focus on
$\langle
{V^{\bar{5}}_{-\frac{1}{2}}}^{*}\,V^{\bar{5}}_{-\frac{1}{2}} \,
{V^{10}_{-\frac{1}{2}}}^{*} V^{10}_{-\frac{1}{2}} \rangle$ and
afterwards we will compute $\langle
{V^{10}_{-\frac{1}{2}}}^{*}\,V^{10}_{-\frac{1}{2}} \,
{V^{10}_{-\frac{1}{2}}}^{*} V^{10}_{-\frac{1}{2}} \rangle$, which
was already examined in \cite{Klebanov:2003my}  \\
\\
\textbf{The amplitude $\langle
{V^{\bar{5}}_{-\frac{1}{2}}}^{*}\,V^{\bar{5}}_{-\frac{1}{2}} \,
{V^{10}_{-\frac{1}{2}}}^{*} V^{10}_{-\frac{1}{2}} \rangle$}\\
\\
We start with the region $-\frac{1}{2}<\theta_{1}<0\, ,
-\frac{1}{2}<\theta_{2}<0 \,,\frac{1}{2}<\theta_{3}<1$ and calculate
the amplitude
\begin{align*}
 \int\, \prod^4_{i=1} \mathrm{d}z_i  \,\,\langle
{V^{\bar{5}}_{-\frac{1}{2}}}^{*}(z_1)\,V^{\bar{5}}_{-\frac{1}{2}}
(z_2) \, {V^{10}_{-\frac{1}{2}}}^{*}(z_3) V^{10}_{-\frac{1}{2}}(z_4)
\rangle\,\,,
\end{align*}
where the vertex operators are in the previous section. Note that
all the vertex operators are in the $(-\frac{1}{2})$-ghost pictures,
which guarantees a total ghost charge of $-2$ on the disk. Plugging
in the vertex operators we see that in order to calculate the
amplitude we need the following correlators
\begin{align}
\begin{gathered}
\Big\langle \prod_{i=1}^{4}  \mathrm{e}^{\mathrm{i} k_i \cdot
X(z_i)}    \Big\rangle = \prod_{\substack{i,j=1\\
i< j}}^{4} {z_{ij}}^{\alpha'\,k_i\cdot k_j} \qquad  \Big\langle
\mathrm{e}^{-\frac{\phi}{2}(z_1)}\,
\mathrm{e}^{-\frac{\phi}{2}(z_2)}\,
\mathrm{e}^{-\frac{\phi}{2}(z_3)}\,
\mathrm{e}^{-\frac{\phi}{2}(z_4)}\,\Big\rangle =\prod_{\substack{i,j=1\\
i< j}}^{4} z^{-\frac{1}{4}}_{ij}\\
\\
 {\bar{u}^{\dot{\alpha}}}_1 \, {u_{\alpha}}_2\,
\bar{u}^{\dot{\beta}}_3 \, u_{\beta\,4}\, \langle
\tilde{S}_{\dot{\alpha}}(z_1) \, S^{\alpha}(z_2)
\tilde{S}_{\dot{\beta}}(z_3) S^{\beta} (z_4)\rangle = \bar{u}_1
\gamma^{\mu} u_2 \,\, \bar{u}_3 \gamma_{\mu} u_4 \,
z^{-\frac{1}{2}}_{13} z^{-\frac{1}{2}}_{24}\,\,, \\
\\
\end{gathered}
\end{align}
where $z_{ij}$ denotes $z_i-z_j$. The correlator involving the four
fermionic twist fields takes an easy form, since we can bosonize the
spin fields
\begin{align}
\Big\langle \prod_{i=1}^{4}  \mathrm{e}^{\mathrm{i}\, \alpha_i
H(z_i)}    \Big\rangle = \prod_{\substack{i,j=1\\i< j}}^{4}
{z_{ij}}^{\alpha_i\cdot \alpha_j}\,\,.
\end{align}
The correlator for the bosonic twist fields is more involved. Using
the stress energy tensor method, the quantum part of four bosonic
twist fields with two independent angles evaluates to
\cite{{Cvetic:2003ch},{Lust:2004cx}}
\begin{align}
\langle \sigma_{1-\theta}(z_1)\, \sigma_{\theta}(z_2) \,
\sigma_{1-\nu}(z_3) \, \sigma_{\nu} (z_4)\rangle =
z_{12}^{-\theta(1-\theta)} z_{34}^{-\nu(1-\nu)} \left(\frac{z_{13}
\, z_{24}}{z_{14} \, z_{23}}\right)^{\frac{1}{2} (\theta + \nu)
-\theta\nu} I^{-\frac{1}{2}}(x) \,\,,
\end{align}
with $x=\frac{z_{12}z_{34}}{z_{13}z_{24}}$ and $I(x)$ is given by
\begin{align*}
I(x)= \frac{1}{2\pi} \big[B_1(\theta,\nu) \, \overline{G_2}(x)
H_1(1-x)+ B_2(\theta,\nu) \, G_1(x) \overline{H_2}(1-x)\big]\,\,,
\end{align*}
where
\begin{align*}
\begin{gathered}
B_1(\theta,\nu)=\frac{\Gamma(\theta)\,\Gamma(1-\nu)}{\Gamma(1+\theta-\nu)}\qquad
B_2(\theta,\nu)=\frac{\Gamma(\nu)\,\Gamma(1-\theta)}{\Gamma(1+\nu-\theta)}\\
\\
G_1(x)= {_2F}_1[\theta,1-\nu,1;x]\qquad
G_2(x)= {_2F}_1[1-\theta,\nu,1;x]\\
\\
H_1(x)= {_2F}_1[\theta,1-\nu,1+\theta-\nu;x]\qquad
 H_2(x)={_2F}_1[1-\theta,\nu,1-\theta+\nu;x]\,\,.
 \end{gathered}
\end{align*}
Applying the correlators, the amplitude becomes
\begin{align*}
\emph{A}&= \mathrm{i} Tr({{\Lambda^{\bar{5}}_1}^{\dagger}
\,\Lambda_2 \,{\Lambda^{10}_3}^{\dagger} \,\Lambda^{10}_4}) \,
\bar{u}_1 \gamma^{\mu} u_2 \, \bar{u}_3 \gamma_{\mu} u_4 (2\pi)^{4}
\delta^{(4)}\left(\sum^4_{i=1}k_i\right)  \\
  \\& \qquad   \times  \int\, \prod^4_{i=1} \mathrm{d} z_i \frac{
 \left[I\left(-\theta_1, 1+\nu_1,
x\right) \,I\left(-\theta_2, 1+\nu_2, x\right) \,I\left(1-\theta_3,
1+\nu_3, x\right)\right]^{-\frac{1}{2}}}
{(z_{12}\,z_{34})^{\alpha'\,s+1}
  \,(z_{13}\,z_{24})^{\alpha'\,t}(z_{14}\,z_{23})^{\alpha'\,u+1}}\,\,,
\end{align*}
where $s$, $t$ and $u$ are the Mandelstam variables
\begin{align*}
s=-(k_1+k_2)^2 \qquad t=-(k_1+k_3)^2 \qquad u=-(k_1+k_4)^2\,.
\end{align*}
The conformal Killing group can be used to fix three of the vertex
operator positions. A convenient choice is
\begin{align*}
z_1=0 \qquad z_2=x \qquad z_3=1 \qquad z_4=z_{\infty}= \infty\,\,,
\end{align*}
which implies the $c$-ghost contribution
\begin{align*}
\langle\, c\,(0)\,\,  c\,(1) \,\,c\,(z_{\infty})\rangle  =
z^2_{\infty}\,\,.
\end{align*}
After fixing three positions, we are left with an integral over one
worldsheet variable
\begin{align*}
    \emph{A}&= \mathrm{i} C_A \,Tr({{\Lambda^{\bar{5}}_1}^{\dagger} \,\Lambda^{\bar{5}}_2
\,{\Lambda^{10}_3}^{\dagger} \,\Lambda^{10}_4}) \, \bar{u}_1
\gamma^{\mu} u_2 \, \bar{u}_3 \gamma_{\mu} u_4   (2\pi)^{4}
\delta^{(4)} \left(\sum^4_{i=1}k_i\right)  \\
 \\
 &\qquad \times\int^1_0\, \mathrm{d}x \,
\frac{ \left[I\left(-\theta_1, 1+\nu_1, x\right) \,I\left(-\theta_2,
1+\nu_2, x\right) \,I\left(1-\theta_3, 1+\nu_3,
x\right)\right]^{-\frac{1}{2}}}{x^{\alpha'\,s+1}
 (1-x)^{\alpha'\,u+1}}.
\end{align*}
In order to obtain the full amplitude we need to sum over all
possible orderings
\begin{align*}
\emph{A}_{total}=& \text{C} \, \Big(
Tr({{\Lambda^{\bar{5}}_1}^{\dagger}\, \Lambda^{\bar{5}}_2
 \,\Lambda^{10}_4\,{\Lambda^{10}_3}^{\dagger}} + Tr({ \Lambda^{\bar{5}}_2 \,
 {\Lambda^{\bar{5}}_1}^{\dagger} \, {\Lambda^{10}_3}^{\dagger}}  \,\Lambda^{10}_4  ) \Big)
 \nonumber
 \int^0_{-\infty} \mathrm{d}x \,U(x)
 \\  & \text{C} \, \Big( Tr({{\Lambda^{\bar{5}}_1}^{\dagger}\, \Lambda^{\bar{5}}_2 \,{\Lambda^{10}_3}^{\dagger}
 \,\Lambda^{10}_4} + Tr({ \Lambda^{\bar{5}}_2 \,
 {\Lambda^{\bar{5}}_1}^{\dagger} \,\Lambda^{10}_4 \, {\Lambda^{10}_3}^{\dagger}}    ) \Big)
 \int^1_{0} \,\, \mathrm{d}x \,U(x)
 \\
\nonumber & \text{C} \, \Big( Tr({{\Lambda^{\bar{5}}_1}^{\dagger}
\,{\Lambda^{10}_3}^{\dagger} \, \Lambda^{\bar{5}}_2
 \,\Lambda^{10}_4} + Tr({
 {\Lambda^{\bar{5}}_1}^{\dagger} \,\Lambda^{10}_4 \,  \Lambda^{\bar{5}}_2 \, {\Lambda^{10}_3}^{\dagger}}    )
 \Big)
 \int^{\infty}_{1} \mathrm{d}x \,U(x)\,\,,
\end{align*}
with
\begin{align*}
\text{C}=\mathrm{i} \,C_A  \, \bar{u}_1 \gamma^{\mu} u_2 \,
\bar{u}_3 \gamma_{\mu} u_4   (2\pi)^{4} \delta^{(4)}
\left(\sum^4_{i=1}k_i\right)
\end{align*}
and
\begin{align*}
U(x)&=  \frac{
 \left[I\left(-\theta_1, 1+\nu_1, x\right) \,I\left(-\theta_2,
1+\nu_2, x\right) \,I\left(1-\theta_3, 1+\nu_3,
x\right)\right]^{-\frac{1}{2}}}{x^{\alpha'\,s+1}
 (1-x)^{\alpha'\,u+1}}\,\,.
\end{align*}
Calculating the traces for the third term by plugging in the
respective Chan-Paton factors immediately shows that they vanish.
Explicit computation of the traces leads to the identities
$Tr({{\Lambda^{\bar{5}}_1}^{\dagger} \,\Lambda^{\bar{5}}_2
\,{\Lambda^{10}_3}^{\dagger}
\,\Lambda^{10}_4})=Tr({\Lambda^{\bar{5}}_2 \,
{\Lambda^{\bar{5}}_1}^{\dagger}
 \,\Lambda^{10}_4 \,{\Lambda^{10}_3}^{\dagger}})$ and
 $Tr({{\Lambda^{\bar{5}}_1}^{\dagger}  \,\Lambda^{\bar{5}}_2
\,\Lambda^{10}_4
\,{\Lambda^{10}_3}^{\dagger}})=Tr({\Lambda^{\bar{5}}_2 \,
{\Lambda^{\bar{5}}_1}^{\dagger} \,{\Lambda^{10}_3}^{\dagger}
\,\Lambda^{10}_4 })$ and thus the amplitude takes the form
\begin{align}
\nonumber A_{total}= 2\mathrm{i} C_A  \, &\bar{u}_1 \gamma^{\mu} u_2
\, \bar{u}_3 \gamma_{\mu} u_4   (2\pi)^{4} \delta^{(4)}
\left(\sum^4_{i=1}k_i\right)\\  &\times \Big(
Tr({{\Lambda^{\bar{5}}_1}^{\dagger}\, \Lambda^{\bar{5}}_2
\,{\Lambda^{10}_3}^{\dagger}
 \,\Lambda^{10}_4}) K(\theta_1,\theta_2,\theta_3)  + Tr({{\Lambda^{\bar{5}}_1}^{\dagger}  \,\Lambda^{\bar{5}}_2
\,\Lambda^{10}_4 \,{\Lambda^{10}_3}^{\dagger}})
T(\theta_1,\theta_2,\theta_3) \Big), \label{final result 551010 1}
\end{align}
with
\begin{align}
K(\theta_1,\theta_2,\theta_3)=\int^1_{0} \,\, \mathrm{d}x \,U(x)
\qquad T(\theta_1,\theta_2,\theta_3)= \int^0_{-\infty} \,\,
\mathrm{d}x \,U(x) \label{definition of K and T} .
\end{align}
In the field theory, the first term corresponds to proton decay via
a gauge boson, while the second one describes the proton decay
mediated via a Higgs particle, arising from the Yukawa interaction
$\mathbf{10}\,\mathbf{\bar{5}}\,\mathbf{\bar{5}_{H}}$.\\
Finally we replace the $\nu$'s by the angles $\theta$
\begin{align*}
\nu_1=2\theta_1 \qquad \nu_2=2\theta_2 \qquad \nu_3=-2+2\theta_3
\end{align*}
and obtain for $U$
\begin{align}
U(x)&= \frac{\left[I\left(-\theta_1, 1+2\theta_1, x\right)
I\left(-\theta_2, 1+2\theta_2, x\right) \,I\left(1-\theta_3,
-1+2\theta_3, x\right)\right]^{-\frac{1}{2}}}{x^{\alpha'\,s+1}
 (1-x)^{\alpha'\,u+1}} \label{K1}\,\,.
\end{align}
Applying the same procedure for the other sector we obtain
\begin{itemize}
\item{$\frac{1}{2}<\theta_{1}<1\qquad \frac{1}{2}<\theta_{2}<1
\qquad \frac{1}{2}<\theta_{3}<1$ \\
\\
The amplitude
\begin{align*}
 \int\, \prod^4_{i=1} \mathrm{d}z_i  \,\,\langle
{V^{\bar{5}}_{-\frac{1}{2}}}^{*}(z_1)\,V^{\bar{5}}_{-\frac{1}{2}}
(z_2) \, {V^{10}_{-\frac{1}{2}}}^{*}(z_3) V^{10}_{-\frac{1}{2}}(z_4)
\rangle\,\,,
\end{align*}
takes the form
\begin{align}
\label{final result 551010 2} \emph{A}_{total}=&  2\mathrm{i}C_A  \,
\bar{u}_1 \gamma^{\mu} u_2 \, \bar{u}_3 \gamma_{\mu} u_4 \,
(2\pi)^{4} \delta^{(4)} \left(\sum^4_{i=1}k_i\right) \\ \nonumber
&\,\,\, \times \Big(Tr({{\Lambda^{\bar{5}}_1}^{\dagger}
\,\Lambda^{\bar{5}}_2 \,{\Lambda^{10}_3}^{\dagger}
\,\Lambda^{10}_4})K(\theta_1,\theta_2,\theta_3) +
Tr({{\Lambda^{\bar{5}}_1}^{\dagger} \,\Lambda^{\bar{5}}_2
\,\Lambda^{10}_4 \,{\Lambda^{10}_3}^{\dagger}})
T(\theta_1,\theta_2,\theta_3) \Big)\,\,,
\end{align}
with $K$ and $T$ defined in \eqref{definition of K and T} and $U$
given by
\begin{align}
U(x)&= x^{-\alpha'\,s-1} (1-x)^{-\alpha'\,u-1}
\prod^3_{I=1}[I(1-\theta_I,-1+2\theta_I,x)]^{-\frac{1}{2}}
\label{K2}\,\,.
\end{align}

}

\end{itemize}
\textbf{The amplitude $\langle
{V^{10}_{-\frac{1}{2}}}^{*}\,V^{10}_{-\frac{1}{2}} \,
{V^{10}_{-\frac{1}{2}}}^{*} V^{10}_{-\frac{1}{2}} \rangle$}\\
\\
Note that in both cases, $\frac{1}{2}<\theta_{1,2}<1\,,
-\frac{1}{2}<\theta_3<0$ and $\frac{1}{2}<\theta_{1,2,3}<1$, the
vertex operators for the matter fields transforming as $\mathbf{10}$
take the same form. Thus the computation of the amplitude $\langle
{V^{10}_{-\frac{1}{2}}}^{*}\,V^{10}_{-\frac{1}{2}} \,
{V^{10}_{-\frac{1}{2}}}^{*} V^{10}_{-\frac{1}{2}} \rangle$ is
identical for both cases. We use the same correlators stated above
except for the one involving the bosonic twist fields, which takes a
simpler form, since it involves only one independent angle
\cite{Cvetic:2003ch}
\begin{align}
\langle \sigma_{1-\theta}(z_1)\, \sigma_{\theta}(z_2) \,
\sigma_{1-\theta}(z_3) \, \sigma_{\theta} (z_4)\rangle =
\left(\frac{z_{13} \, z_{24}}{z_{12}\, z_{14} \, z_{23}
\,z_{34}}\right)^{\theta(1-\theta)}\, L^{-\frac{1}{2}}(x)
\end{align}
with
\begin{align*}
L(x)=\frac{1}{\sin(\pi\,\theta)} \, _2F_1[\theta,1-\theta,1,x] \,
_2F_1[\theta,1-\theta,1,1-x].
\end{align*}
Plugging in all the correlators and fixing three vertex operator
positions we obtain
\begin{align*}
\emph{A}_{total}&= \mathrm{i}\, C_{A}'
\Big(Tr({{\Lambda^{10}_1}^{\dagger} \,\Lambda^{10}_2
\,{\Lambda^{10}_3}^{\dagger} \,\Lambda^{10}_4})+
Tr({{\Lambda^{10}_1}^{\dagger} \,\Lambda^{10}_4
\,{\Lambda^{10}_3}^{\dagger} \,\Lambda^{10}_2})\Big) \, (2\pi)^{4}
\delta^{(4)} \left(\sum^4_{i=1}k_i\right)\\ \nonumber
 &\qquad \qquad \qquad \times \bar{u}_1 \gamma^{\mu} u_2 \, \bar{u}_3
\gamma_{\mu} u_4  \int^1_0 \mathrm{d}x \,  x^{-\alpha'\,s-1}\,
(1-x)^{-\alpha'\,u-1} \prod^3_{I=1} L^{-\frac{1}{2}}(1+\nu_I,x)\,\,.
\end{align*}
Finally we replace the $\nu_I$ by $\theta_I$ and obtain
\begin{itemize}
\item{ $-\frac{1}{2}<\theta_{1}<0\qquad -\frac{1}{2}<\theta_{2}<0
\qquad \frac{1}{2}<\theta_{3}<1$\\
\begin{align}
\nonumber \emph{A}_{total}&= \mathrm{i} C_{A}'
Tr\Big({\Lambda^{10}_1}^{\dagger} \Lambda^{10}_2
{\Lambda^{10}_3}^{\dagger} \Lambda^{10}_4+{\Lambda^{10}_1}^{\dagger}
\Lambda^{10}_4 {\Lambda^{10}_3}^{\dagger} \Lambda^{10}_2\Big)\\
& \qquad \qquad \qquad \times (2\pi)^{4} \delta^{(4)}
\left(\sum^4_{i=1}k_i\right) \bar{u}_1 \gamma^{\mu} u_2 \, \bar{u}_3
\gamma_{\mu} u_4  \,M (\theta_1,\theta_2,\theta_3) \label{final
result 10101010 1}
\end{align}
with
\begin{align}
M(\theta_1,\theta_2,\theta_3)= \int^1_0 \mathrm{d}x \,
\frac{x^{-\alpha'\,s-1}\,
 (1-x)^{-\alpha'\,u-1}}
{ L^{\frac{1}{2}}(1+2\theta_1,x)\,
 L^{\frac{1}{2}}(1+2\theta_2,x)\,
 L^{\frac{1}{2}}(-1+2\theta_3,x)}\,\,.
\label{numerical M1}
\end{align}

} \item{$\frac{1}{2}<\theta_{1}<1\qquad
\frac{1}{2}<\theta_{2}<1 \qquad \frac{1}{2}<\theta_{3}<1$\\
\begin{align}
\nonumber
\emph{A}_{total}&= \mathrm{i} C_{A}'
 Tr \Big( {\Lambda^{10}_1}^{\dagger}
\Lambda^{10}_2 {\Lambda^{10}_3}^{\dagger} \Lambda^{10}_4+
{\Lambda^{10}_1}^{\dagger} \Lambda^{10}_4 {\Lambda^{10}_3}^{\dagger}
\Lambda^{10}_2\Big)  \\
& \qquad \qquad \qquad\times (2\pi)^{4} \delta^{(4)}
\left(\sum^4_{i=1}k_i\right) \bar{u}_1 \gamma^{\mu} u_2 \, \bar{u}_3
\gamma_{\mu} u_4  \,M (\theta_1,\theta_2,\theta_3) \label{final
result 10101010 2}
\end{align}
with
\begin{align}
M(\theta_1,\theta_2,\theta_3)= \int^1_0 \mathrm{d}x \,
\frac{x^{-\alpha'\,s-1}\, (1-x)^{-\alpha'\,u-1}} {
L^{\frac{1}{2}}(2\theta_1-1,x)\,
 L^{\frac{1}{2}}(2\theta_2-1,x)\,
 L^{\frac{1}{2}}(2\theta_3-1,x)}\,\,.
\label{numerical M2}
\end{align}}
\end{itemize}
The $\langle {V^{10}_{-\frac{1}{2}}}^{*}\,V^{10}_{-\frac{1}{2}} \,
{V^{10}_{-\frac{1}{2}}}^{*} V^{10}_{-\frac{1}{2}} \rangle$ does not
involve an Higgs exchange, since couplings of the form
$\mathbf{10}\,\mathbf{10}\,\mathbf{5_H}$ are absent due to the
$U(1)$ charge conversation \cite{Cvetic:2002pj}.
\subsection*{Normalization}
In this section we determine the two undetermined constants $C_A$
and $C^{\prime}_A$ in the string amplitudes computed above. We will
use the fact that even in the low energy limit the integrals
\eqref{definition of K and T}, \eqref{numerical M1} and
\eqref{numerical M2} are convergent in the limit $x \rightarrow 0$,
which corresponds to a gauge boson exchange. Factorizing the
amplitude into two three point functions allows us to normalize it.
We start with the amplitude $\langle
{V^{\bar{5}}_{-\frac{1}{2}}}^{*}\,V^{\bar{5}}_{-\frac{1}{2}} \,
{V^{10}_{-\frac{1}{2}}}^{*} V^{10}_{-\frac{1}{2}} \rangle$ and turn
later to $\langle {V^{10}_{-\frac{1}{2}}}^{*}\,V^{10}_{-\frac{1}{2}}
\, {V^{10}_{-\frac{1}{2}}}^{*} V^{10}_{-\frac{1}{2}} \rangle$.
\\
\\
\textbf{The amplitude $\langle
{V^{\bar{5}}_{-\frac{1}{2}}}^{*}\,V^{\bar{5}}_{-\frac{1}{2}} \,
{V^{10}_{-\frac{1}{2}}}^{*} V^{10}_{-\frac{1}{2}} \rangle$ }\\
\\
We first examine the limit $x \rightarrow 0$ and will see that even
in the low energy limit the integral is convergent, due to the
special
kinematics of this problem.  \\
\\
\emph{Limit $x \rightarrow 0$}
\\
\\
As $x\rightarrow 0$ the hypergeometric functions behave like
\begin{align}
F(a,b,1,x)\rightarrow 1 \qquad F(a,b,a+b,1-x) \rightarrow
\frac{\Gamma(a+b)}{\Gamma(a)\Gamma(b)}\,\,
\ln\left(\frac{\kappa(a,b)}{x}\right) \,\,,\label{asymptotic
behavior1}
\end{align}
with
\begin{align*}
\ln{\kappa(a,b)}=2\psi(1)-\psi(a)-\psi(b)\,\,.
\end{align*}
Applying \eqref{asymptotic behavior1} $I$ takes the form
\begin{align*}
\lim_{x\rightarrow 0} I(\theta,\nu,x) = \frac{1}{\sqrt{\pi}} \,
\ln\left(\frac{\delta(\theta,\nu)}{x}\right)\,\,,
\end{align*}
where $\ln\delta(\theta,\nu)$ is given by
\begin{align*}
\ln\delta(\theta,\nu)=2\psi(1)-\frac{1}{2}\psi(\theta)-
\frac{1}{2}\psi(1-\theta)-\frac{1}{2}\psi(\nu)-\frac{1}{2}\psi(1-\nu)\,\,.
\end{align*}
Therefore even for $s=t=0$ we obtain for the integral \eqref{final
result 551010 1} a convergent expression in the limit $x \rightarrow
0$
\begin{align}
\sim \pi^{3/2}\,  \int_0 \frac{\mathrm{d}x}{x} \,
\ln[1/x]^{-3/2}\,\, .
\end{align}
That allows us to normalize the amplitude by factorizing the
amplitude in the limit $x \rightarrow 0$, where it reduces to a
product of two three-point functions
\begin{align}
A_4 (k_1,k_2,k_3,k_4) = \frac{\mathrm{i}}{2} \int \frac{\mathrm{d}^7
k\,\mathrm{d}^7 k^{'}}{(2\pi)^7} \, \,\frac{\sum_{I J \mu}
A^{I\mu}_j (k_1,k_2,k)A^{I\mu}_j (k_3,k_4,k^{'})
\delta(k-k^{'})}{k^2-\mathrm{i} \epsilon } \label{unitarity}\,\,.
\end{align}
The unusual factor of $\frac{1}{2}$ is introduced to take into
account the doubling in the Chan-Paton factors.\\
The three-point amplitudes describe the exchange of a gauge boson
and are given by
\begin{align}
A^{\mu} (k_1,k_2,k_3)= \mathrm{i} \,g_{D_6} \,(2\pi)^4
\delta^{(4)}\left(\sum^3_{i=1}k_i\right) \bar{u}_1\gamma^{\mu}u_2
Tr({\Lambda^{\bar{5}}_1}^{\dagger} \, \Lambda^{\bar{5}}_2\,
\Lambda_A)\,\,.
\end{align}
Here $\mu$ corresponds to the polarization and $\Lambda_A$ denote
the Chan Paton factors of the gauge boson. The latter takes the form
\begin{align}
\Lambda_A=\left(
\begin{array}{cccc}
\lambda_a & 0&0&0 \\
0& \lambda_a & 0&0 \\
0 & 0&0&0 \\
0 & 0&0&0 \\
\end{array}
 \right)\, ,
\end{align}
where the $\lambda_a$'s are the gauge bosons of $U(5)$ which satisfy
$Tr(\lambda_a\, \lambda_b)=\frac{1}{2} \delta_{ab}$. The
intermediate state is a massless $a-a$ string, which is a gauge
boson, that can carry arbitrary momentum $p$ along the directions of
the D-brane $a$ orthogonal to the intersection. In these directions
we have to integrate over
\begin{align*}
\int \mathrm{d}^{3}q \, \int_0 \mathrm{d}x \,
x^{\alpha'\,q^2-\alpha'\,s-1} = \pi^{3/2} \, (\alpha')^{-3/2} \,
\int_0\mathrm{d}x \, x^{-\alpha'\,s-1} \,[\ln(1/x)]^{-3/2}
\end{align*}
which tells us that the replacement, going from effective field
theory in four dimensions to the form of the string integrand near
$x=0$ is no longer
\begin{align*}
\frac{1}{s} \rightarrow \alpha'\, \int_0 \mathrm{d}x \,
x^{-\alpha'\,s-1}\,\,,
\end{align*}
but
\begin{align}
\int \frac{\mathrm{d}^3 q}{q^2-s} \rightarrow \pi^{3/2}
(\alpha')^{-1/2} \int_0 \mathrm{d}x \, x^{-\alpha' \,
s-1}\,[\ln(1/x)]^{-3/2}\,\, . \label{replacement}
\end{align}
Performing the integral on the right hand side of \eqref{unitarity}
and using the replacement \eqref{replacement} we obtain
\begin{align}
\mathrm{i} \frac{g^2_{D_6} \pi^{5/2} }{2 \alpha'^{1/2}} \,
Tr({{\Lambda^{\bar{5}}_1}^{\dagger} \,\Lambda^{\bar{5}}_2
\,{\Lambda^{10}_3}^{\dagger} \,\Lambda^{10}_4}) \, \bar{u}_1
\gamma^{\mu} u_2 \, \bar{u}_3 \gamma_{\mu} u_4 \,
\delta^{(4)}\left(\sum^4_{i=1}k_i\right) \int_0\, \mathrm{d}x \,
x^{-\alpha'\,s-1} \,[\ln(1/x)]^{-3/2}.
\end{align}
This needs to be the same as \eqref{final result 551010 1} in the
limit $x \rightarrow 0$
\begin{align*}
2\mathrm{i} \, C_A \, Tr({{\Lambda^{\bar{5}}_1}^{\dagger}
\,\Lambda^{\bar{5}}_2 \,{\Lambda^{10}_3}^{\dagger}
\,\Lambda^{10}_4}) \, \bar{u}_1 \gamma^{\mu} u_2 \, \bar{u}_3
\gamma_{\mu} u_4 (2\pi)^{4}\, \pi^{3/2}
\delta^{(4)}\left(\sum^4_{i=1}k_i\right) \int_0\, \mathrm{d}x \,
x^{-s-1} \,[\ln(1/x)]^{-3/2}\,\,,
\end{align*}
which leads us with $g^2_{D_6}= (2\pi)^4 \alpha'^{3/2} g_s$ to the
normalization constant $C_A$
\begin{align}
C_A=\frac{\pi}{2}\,  g_s\,  \alpha' \label{normalization
constant}\,\,.
\end{align}
For the second amplitude one obtains, following the same
procedure, the same normalization constant.\\
\\
\textbf{The amplitude $\langle
{V^{10}_{-\frac{1}{2}}}^{*}\,V^{10}_{-\frac{1}{2}} \,
{V^{10}_{-\frac{1}{2}}}^{*} V^{10}_{-\frac{1}{2}} \rangle$}\\
\\
Note that the amplitude is invariant under the exchange of $x$ and
$1-x$ if one simultaneously interchanges $s$ and $u$. Therefore we
obtain similar limits for $x \rightarrow 0$ and $x \rightarrow 1$.
That is not too surprising taking into account that we expect an
exchange of a gauge boson in both limits.\\
\\
\emph{Limit $ x\rightarrow 0$ and  $ x\rightarrow 1$}\\
\\
Using \eqref{asymptotic behavior1} and taking the low energy limit
$s,t\rightarrow 0$ we get for $x\rightarrow 0$
\begin{align}
\sim \pi^{3/2} \, \int_0 \frac{\mathrm{d}x}{x} \, \ln[1/x]^{-3/2}
\end{align}
and a similar result for $ x\rightarrow 1$
\begin{align}
\sim \pi^{3/2} \, \int^1 \frac{\mathrm{d}x}{1-x} \,
\ln[1/(1-x)]^{-3/2} \,\,.
\end{align}
Following the same procedure as in the case of the amplitude
$\langle
{V^{\bar{5}}_{-\frac{1}{2}}}^{*}\,V^{\bar{5}}_{-\frac{1}{2}} \,
{V^{10}_{-\frac{1}{2}}}^{*} V^{10}_{-\frac{1}{2}} \rangle$ we obtain
for normalization constant ${C_A}'$
\begin{align}
C_{A}'=\pi g_s \, \alpha' \,\,.
\end{align}
\section{Numerical analysis}
We want to compute the contribution of the amplitude which arises
from the four-Fermi interaction in the low energy effective theory.
That means that we take the low energy limit and subtract the $s$,
$t$ and $u$ poles, if present. It turns out that the amplitudes are
divergent only in the limit $x\rightarrow -\infty$. As derived in
appendix B there is no massless exchange in the $u$-channel. The
s-channel requires more explanation, since in general we expect a
massless gauge boson exchange, which leads to an undesired s-pole.
We saw that the integral does not diverge at the s-pole, since we
neglected global effects coming from the internal space. Locally,
the internal dimensions look like a flat space with infinite volume
which leads to a vanishing gauge coupling in four dimensions
\begin{align}
{g_{YM}}^2 \sim \frac{1}{V_{int}}\,\,,
\end{align}
here $V_{int}$ denotes the internal volume and $g_{YM}$ is the gauge
coupling in four dimensions. Thus even if we observe a gauge boson
exchange, we do not see an $s$-pole in our effective low energy
theory. In the limit $x\rightarrow-\infty$, which corresponds to a
$t$-pole, the integral is divergent and in order to obtain the
four-Fermi interaction we have to subtract this pole. A detailed
discussion of the numerical analysis of the integrals $K$,$T$ and
$M$ in the amplitudes \eqref{final result 551010 1},\eqref{final
result 551010 2}, \eqref{final result 10101010 1} and \eqref{final
result 10101010 2} can be found in appendix B, where for
simplification we set $\theta_1=\theta_2=\theta$.
\begin{table}[h]
\begin{center}
\begin{tabular}{|c|c|c|c|c|c|c|c|} \hline
\multicolumn{4}{|p{6.4cm}|}{$-\frac{1}{2}<\theta_{1}<0  \,\,\,
-\frac{1}{2}<\theta_{2}<0 \,\,\, \frac{1}{2}<\theta_{3}<1$} &
\multicolumn{4}{|p{6.4cm}|} {$\frac{1}{2}<\theta_{1}<1\,\,\,\,\,
\frac{1}{2}<\theta_{2}<1 \,\,\,\,\, \frac{1}{2}<\theta_{3}<1$}\\
\hline  $\theta$  & $K$
 & $T$  & $M$
& $\theta$ &$K$& $T$ &
$M$\\
\hline
 \,\,\,\,-.40 \,\,& \,\,\,\,\,\,\,\,\,6.5 \,\,\,\,\,\,\,& \hspace{.6cm}5.4\hspace{.6cm} &
\,\,\,\,\,\,\,10.3\,\,\,\,\,\,& .505 & 1.5 & 1.5 & 2.5
\\\hline
  -.42 & 5.7& 5.1 & 9.4 & .51 & 2.0& 2.1&  3.5
\\\hline
 -.44 &  4.9 & 4.6& 8.3 &    .52  &  2.9 & 2.9 & 4.9
\\\hline
 -.46 & 4.0 &4.0 & 6.9 & .54 & 4.0  & 4.0 & 6.9  \\\hline
 -.48 &  2.9&2.9 & 4.9 &.56 & 4.9  &4.6 & 8.3
\\\hline
-.49 & 2.0 &2.1 & 3.5 & .58 &5.7& 5.1 & 9.4
\\\hline
 -.495 &1.5 &1.5 & 2.5 &
  \,\,\,\,.60 \,\,&  \,\,\,\,\,\,\,\,\, 6.5  \,\,\,\,\,\,\, & \hspace{.6cm}5.4 \hspace{.6cm} &
  \,\,\,\,\,\,\, 10.3\,\,\,\,\,\,  \\ \hline
\end{tabular}
\end{center}
 \caption{
 \label{tab:angles}Contribution to $K$, $T$ and $M$ for
   different angles $\theta$}
\end{table}
Table \ref{tab:angles} shows the contribution $M$ for the string
amplitude $\langle
{V^{10}_{-\frac{1}{2}}}^{*}\,V^{10}_{-\frac{1}{2}} \,
{V^{10}_{-\frac{1}{2}}}^{*} V^{10}_{-\frac{1}{2}} \rangle$ and the
contributions $K$ and $T$ arise from $\langle
{V^{\bar{5}}_{-\frac{1}{2}}}^{*}\,V^{\bar{5}}_{-\frac{1}{2}} \,
{V^{10}_{-\frac{1}{2}}}^{*} V^{10}_{-\frac{1}{2}} \rangle$ for
different angles $\theta$. For $\theta=-1/3$ and $\theta=2/3$  we
observe a second massless fermion which indicates that we now have
$N=2$ supersymmetry. Since our world is chiral we choose $\theta$ in
the ranges, given in table \ref{tab:angles}. 
\\
Note also, that going from the first sector
$-\frac{1}{2}<\theta_{1}<0 \,, -\frac{1}{2}<\theta_{2}<0 \,,
\frac{1}{2}<\theta_{3}<1$ to the second one
$\frac{1}{2}<\theta_{1}<1\,, \frac{1}{2}<\theta_{2}<1\,,
\frac{1}{2}<\theta_{3}<1$ and replacing $\theta$ by $1-\theta$,
simultaneously leads to the same results for $K$, $T$ and $M$. This
is not too surprising, since the respective vertex operators
correspond to the same states if you interchange $\theta$ with
$1-\theta$.
\section{Comparison to Four-Dimensional Field theory}
In this section we want to compare the amplitude obtained due to
massive string states in string theory with the amplitude on the
field theory side. Therefore, we would like to replace all the
string theory parameters such as the string coupling $g_s$ or the
gauge coupling $g_{D6}$ by appropriate expressions using quantities
about which we have some knowledge of, such as
$M_{GUT}$ and $\alpha_{GUT}$. We follow closely the analysis of \cite{Klebanov:2003my}.\\
The action for the gauge fields living on the $D6$-branes is
\begin{align*}
\frac{1}{4g^2_{D_6}}\int \mathrm{d}^7 x \, \text{Tr}
F_{ij}F^{ij}\,\,,
\end{align*}
where the $F_{ij}$ is the Yang-Mills field strength and Tr denotes
the trace in the fundamental representation of $U(N)$. After
compactification on $R \times Q $ the action becomes
\begin{align*}
\frac{V_{Q}}{4g^2_{D_6}}\int \mathrm{d}^4 x \, \text{Tr}
F_{ij}F^{ij},
\end{align*}
where $V_{Q}$ is the volume of $Q$. Keeping in mind the usual
convention $Tr (Q_a Q_b)=\frac{1}{2}\delta_{ab}$ we finally obtain
for the action
\begin{align}
\frac{V_{Q}}{8g^2_{D_6}}\int \mathrm{d}^4 x \, \text{Tr}
F_{ij}F^{ij}. \label{D-branes action}
\end{align}
On the other hand, the GUT action is given by
\begin{align}
\frac{1}{4g^2_{GUT}}\int \mathrm{d}^4 x \, \text{Tr}
F_{ij}F^{ij}\,\,, \label{GUT action}
\end{align}
where $g_{GUT}$ is the GUT coupling. Comparing \eqref{D-branes
action} and \eqref{GUT action}, along with $g^2_{D_6}= (2\pi)^4 g_s
\alpha'^{3/2}$ \cite{Polchinski:1998rr} and
$\alpha_{_{GUT}}=g^2_{_{GUT}}/(4\pi)$, leads to the identification
\begin{align}
\alpha'=\left( \frac{  \alpha_{_{GUT}} \, V_{Q}}{(2\pi)^3 \,
g_s}\right)^{2/3}\,\,.
\end{align}
The volume $V_Q$ enters into the running of the $SU(3)\times
SU(2)\times U(1)$ gauge coupling from high energies to low energies.
Approximately, one can say that $V^{-1/3}_{Q}$ plays the role of the
mass scale unification $M_{GUT}$ in four dimensions. In order to
obtain the exact relation between them one needs to compute the one
loop threshold correction to the gauge coupling, which was done for
M-theory on a manifold of $G_2$ holonomy
\cite{Friedmann:2002ty}\footnote{An explicit computation for the one
loop threshold correction in type IIA string theory was performed in
\cite{Lust:2003ky}, which leads in the limit $g_s\rightarrow1$ to an
equivalent relation.}
\begin{align}
V_{Q} = \frac{L(Q)}{M^3_{GUT}}\,\,,
\end{align}
where $L(Q)$ is a topological invariant, the Ray-Singer torsion. In
\cite{Klebanov:2003my} it is argued that this relation holds true in
Type IIA string theory and thus we finally obtain
\begin{align}
\alpha'=\left( \frac{  \alpha_{_{GUT}} \, L(Q)}{(2\pi)^3 \, g_s\,
M^3_{GUT}}\right)^{2/3}\,\,. \label{string theory  field theory}
\end{align}
We would like to replace all the string parameters in the amplitudes
\eqref{final result 551010 1} and \eqref{final result 551010 2} in
terms of four dimensional field theory quantities. Unfortunately,
equation \eqref{string theory  field theory} still includes two
string parameters $L(Q)$ and $g_s$. The Ray-Singer torsion $L(Q)$
depends crucially on the compact space and takes for simple lens
spaces values around $8$ \cite{Friedmann:2002ty}. In order to
neglect higher order loop amplitudes the string coupling $g_s$ is
better smaller than $1$. On the other hand we are interested in the
largest possible contribution to the enhancement and set therefore
$g_s$
approximately to 1.\\
\\
\emph{Field theory amplitude}
\\
\\
After relating the string parameters to four dimensional field
theory constants, of which we have some experimental knowledge, we
now recall the analysis of proton decay in the $SU(5)$ GUT
model\footnote{As done usually we neglect because of the weakness of
the Yukawa couplings to light fermions the Higgs mediated
Proton decay.}. This treatment closely follows \cite{Langacker:1980js}.\\
The kinetic energy for an $SU(5)$ gauge theory, involving the gauge
field $A$, the fermionic field $\psi_{\bar{5}}$, which transforms as
$\mathbf{\bar{5}}$, and the fermionic field $\psi_{10}$ transforming
as \textbf{10} under the $SU(5)$ takes the form
\begin{align}
T=\frac{1}{4 g^2_{_{GUT}}} Tr(F^2(A))  + \mathrm{i}
\bar{\psi}_{\bar{5}} \gamma^{\mu} D_{\mu} \psi_{\bar{5}} +
\mathrm{i} \bar{\psi}_{10} \gamma^{\mu} D'_{\mu} \psi_{10}
\label{kinetic energy}
\end{align}
with
\begin{align*}
D_{\mu}{\psi_{\bar{5}}}^a = \partial_{\mu} \, {\psi_{\bar{5}}}^{a}
-\frac{\mathrm{i}\, g_{_{GUT}} }{\sqrt{2}} \, {(A_{\mu})}^{a}_{b} \,
{\psi_{\bar{5}}}^b
\end{align*}
and
\begin{align*}
D'_{\mu}{\psi_{10}}^{ab} = \partial_{\mu} \, {\psi_{10}}^{ab}
-\frac{\mathrm{i}\, g_{_{GUT}} }{\sqrt{2}} \, {(A_{\mu})}^{a}_{c} \,
{\psi_{10}}^{cb} -\frac{\mathrm{i}\, g_{_{GUT}} }{\sqrt{2}} \,
{(A_{\mu})}^{b}_{d} \, {\psi_{10}}^{ad}\,\,.
\end{align*}
By explicitly using the antisymmetry of $\psi_{10}$, the latter can
be simplified to
\begin{align*}
D'_{\mu}{\psi_{10}}^{ab} = \partial_{\mu} \, {\psi_{10}}^{ab}
-\frac{\mathrm{i}\, 2g_{_{GUT}} }{\sqrt{2}} \, {(A_{\mu})}^{a}_{c}
\, {\psi_{10}}^{cb}.
\end{align*}
The gauge field $A$ can be displayed as a $5 \times 5$ matrix
\begin{align*}
A_{\mu}= \left(
\begin{array}{cccccc}
& & & |& {X^C_1}_{\mu}& {Y^C_1}_{\mu}  \\
& \frac{1}{\sqrt{2}}\sum_{a} G^a_{\mu} \lambda^a & &|&
{X^C_2}_{\mu} & {Y^C_2}_{\mu}\\
& & &|& {X^C_3}_{\mu}& {Y^C_3}_{\mu}  \\
---& ---& --- & + & ---& ---\\
{X_1}_{\mu} & {X_2}_{\mu} & {X_3}_{\mu}&  & \frac{W^{3}_{\mu}}{\sqrt{2}} & W^+_{\mu} \\
{Y_1}_{\mu}& {Y_2}_{\mu}& {Y_3}_{\mu}& |&   W^-_{\mu}& -\frac{W^{3}_{\mu}}{\sqrt{2}}\\
\end{array} \right) + \frac{B_{\mu}}{\sqrt{30}}
\left(
\begin{array}{ccccc}
-2&&&&\\
&-2&&&\\
&&-2&& \\
&&&3&\\
&&&&3
\end{array}
\right)\,\,,
\end{align*}
where the $\lambda^a$ are the Gell-Mann matrices, the $G^a_{\mu}$
denote the gluon fields of $SU(3)$ and $W^{+}_{\mu}$, $W^{-}_{\mu}$,
$W^{3}_{\mu}$, $B_{\mu}$ are the bosons of the $SU(2)\times U(1)$.
The $X$ and $Y$ are the new gauge bosons that are contained in
$SU(5)$ and do not occur in the standard model. The exchange of
these new gauge bosons leads to Baryon-Lepton
number violating processes and therefore allows proton decay. \\
To make contact to the standard model the $SU(5)$ needs to be
broken, which will be achieved by giving the Higgs field, which
transforms under the $24$-dimensional adjoint representation of
$SU(5)$ an expectation value. This generates a mass $M_X$ of order
of $10^{16}$ Gev for the gauge bosons $X$ and $Y$.\\
From \eqref{kinetic energy} one can easily deduce the effective
four-Fermi interactions which lead to proton decay. Ignoring mixing
effects as well as second and third families one obtains for the
\begin{align}
L_{eff}= \frac{g^2_{_{GUT}}}{2 \,M^2_X} \Big(
\varepsilon_{\alpha\,\beta\,\gamma} \,\bar{u}^{C\gamma}_L\,
\gamma^{\mu} \,u^{\beta}_{L}  \Big) \, \Big( 2 \bar{e}^{+}_L\,
\gamma_{\mu} \, d^{\alpha}_L  + \bar{e}^{+}_R \, \gamma_{\mu} \,
d^{\alpha}_R  \Big)\,\,, \label{four fermi interaction}
\end{align}
where the first factor arises from a
$\mathbf{10}^*\,\mathbf{10}\,\mathbf{10}^*\,\mathbf{10}$ interaction
and the second factor from a
$\mathbf{\bar{5}}^*\,\mathbf{\bar{5}}\,\mathbf{10}^*\,\mathbf{10}$
interaction .\\
\\
\emph{Comparison}
\\
\\
This result \eqref{four fermi interaction} we want to compare with
the string theory contribution. In order to do that we turn on
Wilson lines, that break the $SU(5)$ gauge group into the standard
model ones. Assuming such a mechanism of symmetry breaking exist we
compute the traces of \eqref{final result 551010 1} and \eqref{final
result 10101010 1} only for entries which lead to proton decay. One
obtains for \eqref{final result 551010 1} and \eqref{final result
551010 2}
\begin{align}
A^{\bar{5}\bar{5}1010}_{total}= \mathrm{i} \, (2\pi)^{4}
\delta^{(4)} \left(\sum^4_{i=1}k_i\right) \,\pi g_s \alpha' \Big(
\varepsilon_{\alpha\,\beta\,\gamma} \,\bar{u}^{C\gamma}_L\,
\gamma^{\mu} \,u^{\beta}_{L}  \Big) \, \Big(  \bar{e}^{+}_R \,
\gamma_{\mu} \, d^{\alpha}_R  \Big) \left(K(\theta)
+T(\theta)\right)
\end{align}
and for \eqref{final result 10101010 1} and \eqref{final result
10101010 2}
\begin{align}
\emph{A}^{10101010}_{total}&= 2 \mathrm{i}\, (2\pi)^{4} \delta^{(4)}
\left(\sum^4_{i=1}k_i\right) \, \pi g_s \, \alpha'
  \Big(
\varepsilon_{\alpha\,\beta\,\gamma} \,\bar{u}^{C\gamma}_L\,
\gamma^{\mu} \,u^{\beta}_{L}  \Big) \, \Big(  \bar{e}^{+}_L\,
\gamma_{\mu} \, d^{\alpha}_L \Big) \, M(\theta)\,.
\end{align}
Comparing the string theory proton decay rate with the one from four
dimensional gauge theory one obtains
\begin{align}
\frac{\Gamma_{ST}(p\rightarrow\pi^{0}\,
e^{+})}{\Gamma_{FT}(p\rightarrow\pi^{0}\, e^{+})} =
\left(\frac{g^{1/3}_s \, L(Q)^{2/3}}{8 \pi^2 \alpha^{1/3}_{_{GUT}}}
\right)^2 \, \left(\frac{M_X}{M_{_{GUT}}}\right)^{4} \,
\left(\frac{(K+T)^2+ 4M^2}{4}\right)\,. \label{ratio}
\end{align}
Most recent calculations \cite{Hisano:2000dg} for the proton decay mediated
via gauge bosons in an $SU(5)$-GUT model gave the lifetime $\tau^{FT}_{p}$ in
terms of gauge boson mass $M_X$ and $\alpha_{_{GUT}}$
\begin{align}
\tau^{FT}_{p} = 1.6 \times 10^{36} years
\left(\frac{0.04}{\alpha_{_{GUT}}}\right)^2
\left(\frac{M_X}{20^{16}GeV}\right)^{4}\,\,. \label{lifetime field
theory}
\end{align}
This leads with the values $M_X=M_{_{GUT}}=2\times10^{16} GeV$ and
$\alpha_{_{GUT}}=0.04$ to a proton lifetime of $1.6 \times 10^{36}
years$. The present lower bound on the proton lifetime for $p
\rightarrow \pi^{0} \, e^{+}$ is $1.6\times 10^{33} years$
\cite{Eidelman:2004wy} and even the next generation proton decay
experiments, based on underground water Cherenkov detectors will
reach a lower bound not larger than $10^{35}years$
\cite{Jung:1999jq}. Therefore in the near future, unless there is an
enhancement to the proton decay amplitude, we will not observe the
proton decay via gauge boson exchange. Using \eqref{ratio} and
\eqref{lifetime field theory} the proton lifetime in the considered
type IIA string models is
\begin{align}
\tau^{ST}_p   \approx 1.6\times 10^{36} years \frac{54^2}{L^{4/3}(Q)
\, g^{2/3}_s\,((K+T)^2+ 4M^2)}
\left(\frac{0.04}{\alpha_{GUT}}\right)^{4/3}
\left(\frac{M_{_{GUT}}}{20^{16}GeV}\right)^{4}\,\,, \label{proton
lifetime string}
\end{align}
where $\frac{L^{4/3}(Q) \, g^{2/3}_s\,((K+T)^2+ 4M^2)}{54^2}$ is the
string enhancement factor. Note that in \eqref{proton lifetime
string} the heavy gauge boson mass $M_X$, which is model dependent,
is absent and the proton lifetime depends only on $M_{_{GUT}}$. We
also observe an anomalous power of $\alpha_{_{GUT}}$ in
\eqref{proton lifetime string}
indicating the stringy nature of the enhancement.\\
Let us examine the enhancement factor $\frac{L^{4/3}(Q) \,
g^{2/3}_s\,((K+T)^2+ 4M^2)}{54^2}$. As already mentioned earlier the
Ray-Singer torsion is around $8$ for lens spaces with small
fundamental group. The string coupling takes values between $0$ and
$1$, but in order to obtain the largest possible enhancement to the
proton decay amplitude we assume it is approximately $1$. Table
\ref{tab:angles} shows that $M$ ranges between $5$ and $10$, while
$K+T\approx 1.2\times M$, leading with the numerical
four-dimensional $SU(5)$ supersymmetric values
$M_{_{GUT}}=2\times10^{16} GeV$ and $\alpha_{_{GUT}}=0.04$ to a
proton lifetime $\tau^{ST}_p =(0.5-2.1)\times 10^{36} years$. We see
that although there is in addition to the contribution to the
four-Fermi interaction which in field theory are due to gauge boson
exchange, there is also a contribution due to terms that in field
theory arise from Higgs particle exchange, the total string
contribution is not large enough to lead to a considerable
enhancement in the proton decay rate.\\
The dimension six operators $ \mathbf{\bar{5}^{^{*}}}
\mathbf{\bar{5}} \mathbf{10}^{^{*}} \mathbf{10} $ have in contrast
to the operators $ \mathbf{10^{^{*}}} \mathbf{10} \mathbf{10}^{^{*}}
\mathbf{10}$ a second proton decay mode; they lead in addition to
the decay mode $p \rightarrow \pi^0\,e^+$ also to $p\rightarrow
\pi^+ \bar{\nu}$. Plugging in the respective entries in \eqref{final
result 551010 1} leading to the mode $p\rightarrow \pi^+ \bar{\nu}$
one obtains
\begin{align}
A^{\bar{5}\bar{5}1010}_{total}= \mathrm{i} \, (2\pi)^{4}
\delta^{(4)} \left(\sum^4_{i=1}k_i\right) \,\pi g_s \alpha' \Big(
\varepsilon_{\alpha\,\beta\,\gamma} \,\bar{u}^{C\gamma}_L\,
\gamma^{\mu} \,d^{\beta}_{L}  \Big) \, \Big(  \bar{\nu}^{C}_R \,
\gamma_{\mu} \, d^{\alpha}_R  \Big) \left(K(\theta)
+T(\theta)\right)\,\,.
\end{align}
Within  the field theory the effective interaction
\begin{align}
L_{eff}= \frac{g^2_{_{GUT}}}{2 \,M^2_X} \Big(
\varepsilon_{\alpha\,\beta\,\gamma} \,\bar{u}^{C\gamma}_L\,
\gamma^{\mu} \,d^{\beta}_{L}  \Big) \, \Big(  \bar{\nu}^{C}_R \,
\gamma_{\mu} \, d^{\alpha}_R  \Big)\,\,, \label{four fermi
interaction 1}
\end{align}
the ratio between the proton decay rates is given by
\begin{align}
\frac{\Gamma_{ST}(p\rightarrow\pi^{+}\,
\bar{\nu})}{\Gamma_{FT}(p\rightarrow\pi^{+}\, \bar{\nu})} =
\left(\frac{g^{1/3}_s \, L(Q)^{2/3}}{16 \pi^2 \alpha^{1/3}_{_{GUT}}}
\right)^2 \, \left(\frac{M_X}{M_{_{GUT}}}\right)^{4} \,
\Big(K+T\Big)^2\,\,. \label{ratio 1}
\end{align}
For this decay mode the string enhancement to the proton decay rate
is even smaller than for the mode $p \rightarrow \pi^0\,e^+$ due to
the absence of the $ \mathbf{10^{^{*}}} \mathbf{10}
\mathbf{10}^{^{*}} \mathbf{10}$ interaction term. For the same
choice of parameter as above (in addition we assume that
$M_X=M_{GUT}$) the ratio \eqref{ratio 1} takes values between $0.2$
and $0.8$.
\section{Conclusions}
In this paper we computed the local, string contribution to the
proton decay rate for supersymmetric SU(5) GUT's based on
intersecting D6-brane constructions in Type IIA string theory
orientifolds by explicitly calculating the string amplitude
contribution to the dimension six operators. If the compactification
volume is larger than the string scale, world-sheet instanton
effects are negligible and the local contribution is the dominant
one. In the computation presented, we assumed that the matter fields
$\mathbf{\bar{5}}$ and $\mathbf{10}$ are located at the same
intersections on top of each other,  and thus the leading string
amplitude contributions have no suppressions from area factors. In
this case the amplitudes give the largest possible contribution to
the proton decay rate. In contrast to the authors
\cite{Klebanov:2003my}, who only considered the amplitude $\langle
\mathbf{10^{^{*}}} \mathbf{10} \mathbf{10}^{^{*}} \mathbf{10}
\rangle $, we also included the explicit calculation of the string
amplitude  for $ \mathbf{\bar{5}^{^{*}}} \mathbf{\bar{5}}
\mathbf{10}^{^{*}} \mathbf{10} $ operators.  \\
As a by-product we explicitly constructed the vertex operators for
any massless string excitation at supersymmetric D-brane
intersections arising in Type IIA toroidal orientifolds.
Specifically, by employing explicit string vertex operators for the
$\mathbf{10}$ and $\mathbf{\bar{5}}$ chiral superfields, we
calculated explicitly string theory amplitudes contributing to the
proton decay via dimension six operators. In the analysis we chose
the most symmetric configurations in order to maximize proton decay
rates for the above dimension six operators and we obtain a small
enhancement relative to the field theory result. In contrast to the
string amplitude $\langle \mathbf{10^{^{*}}} \mathbf{10}
\mathbf{10}^{^{*}} \mathbf{10} \rangle $, where only the gauge boson
exchange contributes to the proton decay rate for the amplitude
$\langle \mathbf{\bar{5}^{^{*}}} \mathbf{\bar{5}} \mathbf{10}^{^{*}}
\mathbf{10}\rangle $ there is an additional contribution
corresponding to the proton decay mediated via Higgs particle.
\\
After relating the string theory result to the field theory
computations we obtain for the proton lifetime in type IIA string
theory models
\begin{align}
\tau^{ST}_p   \approx 1.6\times 10^{36} years \frac{54^2}{L^{4/3}(Q)
\, g^{2/3}_s\,((K+T)^2+ 4M^2)}
\left(\frac{.04}{\alpha_{GUT}}\right)^{4/3}
\left(\frac{M_{_{GUT}}}{20^{16}GeV}\right)^{4}\,\,, \label{proton
lifetime string1}
\end{align}
which has an anomalous power of $\alpha_{GUT}$ indicating the string
effects. The string enhancement factor depends on the Ray-Singer
torsion, the string coupling $g_s$ and the numerical quantities $M$,
$K$ and $T$. Here the quantity $M$ corresponds to the contribution
arising from the string amplitude $\langle \mathbf{10^{^{*}}}
\mathbf{10} \mathbf{10}^{^{*}} \mathbf{10} \rangle $, while the sum
$K+T$ originates from the string amplitude $\langle
\mathbf{\bar{5}^{^{*}}} \mathbf{\bar{5}} \mathbf{10}^{^{*}}
\mathbf{10}\rangle $, where $K$ is the contribution due to the gauge
boson exchange and $T$ describes the contribution due to the Higgs
particle exchange.
Choosing common values for $L(Q)$, assuming that the string coupling
$g_s$ is approximately $1$ and plugging in the computed numerical
quantities $K$, $M$ and $T$ (see table \ref{tab:angles}) the proton
lifetime \eqref{proton lifetime string1} is $\tau^{ST}_p
=(0.5-2.1)\times 10^{36} years$, and could lead up to a factor of
three shorter lifetime than
that predicted in field theory.\\
\\
\\
\section*{Acknowledgements} We would like to thank Carlo Angelantonj, Andre Brown, Peng Gao,  Paul
Langacker, Tao Liu and Stephan Stieberger for useful discussions. The work  is
supported by an DOE grant DE-FG03-95ER40917 and by the Fay R. and Eugene L.
Langberg Chair.
\newpage
\appendix

\section{Vertex operators for intersecting D-branes}
This appendix discusses the vertex operators of bosonic and
fermionic string states arising in intersecting D-branes based on
the example of intersecting D6-branes. In the following we will
consider D6-branes in flat, non-compact Minkowski space that fill
out the first four dimensions (our actual spacetime) and intersect
in the 3rd, 4th and 5th complex plane. Strings that are stretched
between these D-branes have to satisfy special boundary conditions
in the internal dimensions which leads to non-integer mode
expansions for the degrees of freedom. In the vertex operators for
the corresponding string configuration on introduces bosonic and
fermionic twist fields to take into account these non-integer mode
excitations. These twist fields depend crucially on the choice of
intersecting angles. In this section we will present a instruction
to construct the vertex operators arising from strings stretched
between intersecting D-branes in the NS-sector as well as in the
R-sector.\\
As a first step we deduce the mode expansions for the bosonic and
fermionic degrees of freedom. We start with the NS-sector, where
strings stretched between the intersecting D-branes correspond to
massive scalars in the four-dimensional space-time. After deriving
the mode expansions we quantize the string, impose the condition for
physical states, and obtain the mass formula. Later we will also
deal with strings in the R-sector and show that in this sector we
always have a massless fermion, independent of the choice of the
intersection angles, while in the NS-sector the scalars become
massless only for particular choices of angles that match with the
supersymmetry condition. To get an idea of how the vertex operators
look like, in particular in the internal dimensions, we examine the
operator product expansions (OPE's) of the bosonic and fermionic
fields with specific string excitations. These OPE's show the same
behavior as the OPE's of the twist fields in orbifold theories
\cite{Dixon:1986qv}. Therefore the vertex operators for strings
stretched between intersecting D-branes will involve bosonic and
fermionic twist fields, $\sigma_{\theta}$ and $s_{\theta}$ in the
internal dimensions. The exact knowledge of the OPE's of the bosonic
and fermionic fields with the string states allows us to write the
vertex operators for the string
states in arbitrary intersecting D-brane configurations.\\
An open string stretched between two D-branes at an angle
$\pi\theta_I$ has to fulfill the boundary conditions
\cite{{Arfaei:1996rg},{Abel:2003vv}}
\begin{align}
\begin{gathered}
\partial_{\sigma} X^{p}(\tau,0) = 0 = X^{p+1}(\tau,0) \\
\partial_{\sigma} X_p (\tau,\pi) + \tan{(\pi\theta_I)} \,\,\partial_{\sigma} X_{p+1}(\tau,\pi)=0
\\
 X_{p+1} (\tau,\pi)- \tan{(\pi\theta_I)} \,\, X_p (\tau,\pi)=0\,\,.
\end{gathered}
\end{align}
Given these boundary conditions, we can deduce the mode expansion
for $Z^{I}$ (we use complex coordinates $Z^{I}=X^{2I+2} + \mathrm{i}
X^{2I+3}$) to
\begin{equation}
\left.
\begin{aligned}
Z^{I} (z , \bar{z}) &= \sum_{n}
\frac{\alpha^I_{n-\theta_I}}{(n-\theta_I)} \,\,z^{-n+\theta_I} +
\sum_{n} \frac{\alpha^I_{n+\theta_I}}{(n+\theta_I)}
\,\,\bar{z}^{-n-\theta_I}
 \\
\\
\bar{Z}^{I} (z , \bar{z}) &= \sum_{n}
\frac{\alpha^I_{n+\theta_I}}{(n+\theta_I)} \,\,z^{-n-\theta_I} +
\sum_{n} \frac{\alpha^I_{n-\theta_I}}{(n-\theta_I)}
\,\,\bar{z}^{-n+\theta_I}
\end{aligned}\right\} \text{for}\,\, I=1,2,3  \,\,.
\label{modeexpansionintersecting bosonic}
\end{equation}
Upon quantization the only nonvanishing commutator is
\begin{align*}
[\alpha^{I}_{n \pm \theta}, \alpha^{I'}_{m \mp \theta} ] &= \pm m
\,\delta_{n+m}\, \delta^{II'}.
\end{align*}
World-sheet supersymmetry
\begin{align*}
\delta X^p= \bar{\epsilon} \psi^p
\end{align*}
leads to the same modding for the complexified worldsheet fermions
(here we already used the doubling trick)
\begin{align}
\Psi^I(z)= \sum_{n+\frac{1}{2}} \,\, \psi_{r-\theta_I}^I\,
z^{-r-\frac{1}{2}+\theta_I} \qquad \bar{\Psi}^I(z)=
\sum_{n+\frac{1}{2}} \,\, \psi_{r+\theta_I}^I\,
\bar{z}^{-r-\frac{1}{2}-\theta_I}\,\,\,\,.
\label{modeexpansionintersecting fermionic}
\end{align}
Notice that we consider the NS-sector where the fermions are half
integer modded. The only nonvanishing anti-commutator is given by
\begin{align*}
\{\psi_{m-\theta_I}^I, \psi_{n+\theta_I}^I \}=-\delta_{m,n}\,\,.
\end{align*}
For positve $\theta_I$ ($0<\theta_I<1$) the vacuum in the internal
dimensions is defined by
\begin{equation}
\begin{aligned}
\alpha^I_{m-\theta_I} |\,0\rangle &=0   \qquad m \geq 1 \qquad
\qquad
\psi^I_{r+\theta_I} |\,0\rangle =0 \qquad r \geq \frac{1}{2}\\
 \alpha^I_{m+\theta_I} |\,0\rangle &=0 \qquad m \geq 0\qquad
\qquad \psi^I_{r+\theta_I} |\,0\rangle =0 \qquad r \geq
\frac{1}{2}\,\,.
\end{aligned}
\end{equation}
The physical state constraint requires annihilation with all the
positive modes of the Virasoro generators $L_n$, in particular with
$L_0$, which takes the form
\begin{equation}
\begin{aligned}
L_0&=  \sum^3_{\mu=0} \left\{\sum_{n \,\epsilon \emph{Z}}:
\alpha^{\mu}_{-n} \, \alpha^{\mu}_{n}:+ \sum_{n \,\epsilon \emph{Z}}
n\,:\psi^{\mu}_{-n}\, \psi^{\mu}_{n}:\right\} \\ &+ \sum^3_{I=1}
\left\{\sum_{m\epsilon
\emph{Z}}:\alpha^I_{-m+\theta_I}\alpha^I_{m-\theta_I}: +
\sum_{m\epsilon \emph{Z}}
(m-\theta_I):\psi^I_{-m+\theta_I}\psi^I_{m-\theta_I}:\right\}
+\epsilon_0 \,\,.
\end{aligned}
\label{Virasoro}
\end{equation}
Here $\alpha^{\mu}_n$ and $\psi^{\mu}_n$ denote the excitations in
space-time and $\epsilon_0$ is the zero point energy. Using the fact
that the zero mode $\alpha^{\mu}_0$ represents the momentum of the
string we manipulate equation \eqref{Virasoro} and obtain a mass
formula for the open string in the twisted sector
\begin{equation}
\begin{aligned}
M^2&=\sum^3_{\mu=0} \left\{\sum_{n \epsilon \emph{Z} n}:
\alpha^{\mu}_{-n} \, \alpha^{\mu}_{n}:+ \sum_{n \epsilon \emph{Z}}
n\,:\psi^{\mu}_{-n}\, \psi^{\mu}_{n}:\right\} \\ &+ \sum^3_{I=1}
\left\{\sum_{m\epsilon
 \emph{Z}}:\alpha^I_{-m+\theta_I}\alpha^I_{m-\theta_I}: +
\sum_{m\epsilon \emph{Z}}
(m-\theta_I):\psi^I_{-m+\theta_I}\psi^I_{m-\theta_I}:\right\}
+\epsilon_0 \,\,.
\end{aligned}
\end{equation}
The zero point energy can be computed from the $\zeta$-function
regularization, as we demonstrate in the following (for one internal
dimension only)
\begin{equation}
\begin{aligned}
\epsilon^{I}_0&= \sum^0_{m=-\infty} \left[\alpha_{-m+\theta_I},
\alpha_{m-\theta_I} \right] + \sum^{-1/2}_{m=-\infty}
(r-\theta_I)\left\{\psi_{-r+\theta_I}  , \psi_{r-\theta_I}\right\}\\
&=\zeta[-1,\theta_I]-\zeta[-1,1/2+\theta_I] = -\frac{1}{8}+
\frac{1}{2}\, \theta_I \,\,.
\end{aligned}
\end{equation}
To get an expression for the vertex operators we need to determine
the OPE's of $\Psi^{I}$ and $\bar{\Psi}^{I}$ with some particular
excitations. First we examine the vacuum state $|\,0\rangle$
\begin{align*}
\Psi^{I}(z) \, |\,0\rangle &= \sum^{\infty}_{r=-\infty}
z^{-r-\frac{1}{2}+\theta_I} \psi_{r-\theta_I} |\,0\rangle
=\sum^{-\frac{1}{2}}_{r=-\infty} z^{-r-\frac{1}{2}+\theta_I}
\psi_{r-\theta_I} |\,0\rangle \rightarrow z^{\theta_I} \,
t_{I}(0)\,\,,
\end{align*}
where $t_{I}(0)$ denotes the excited twist field at the
intersection. Similarly we obtain for $\bar{\Psi}^{I}(z)\,
|\,0\rangle$
\begin{align*}
\bar{\Psi}^{I}(z)\, |\,0\rangle \, \rightarrow \,\,z^{-\theta_I} \,
t_{I}'(0)\,\,.
\end{align*}
Using the same procedure, the OPE of $\Psi$ and $\bar{\Psi}$ with
the state $\psi_{-\frac{1}{2}+\theta_I}\, |\,0\rangle$ is
\begin{align*}
&\Psi^{I}(z)\, \psi_{-\frac{1}{2}+\theta_I}\,|\,0\rangle \,
\rightarrow \,\, z^{\theta_I-1} \, t_I(0) \qquad \qquad
\bar{\Psi}^{I}(z)\, \psi_{-\frac{1}{2}+\theta_I}\,|\,0\rangle \,
\rightarrow \,\,z^{1-\theta_I} \, t_I'(0)\,\,.
\end{align*}
Considering a negative  angle $\theta_I$ ($-1<\theta_I<0$) leads to
a different definition of the vacuum
\begin{equation}
\begin{aligned}
\alpha^I_{m-\theta_I} |\,0\rangle &=0   \qquad m \geq 0 \qquad
\qquad
\psi^I_{r-\theta_I} |\,0\rangle =0 \qquad r \geq \frac{1}{2} \\
\alpha^I_{m+\theta_I} |\,0\rangle &=0 \qquad m \geq 1 \qquad \qquad
 \psi^I_{r+\theta_I} |\,0\rangle =0 \qquad r \geq \frac{1}{2}
\end{aligned}
\end{equation}
and the zero point energy, calculated in the same way as above,
takes the form
\begin{align}
\epsilon^{I}_0&=  -\frac{1}{8}- \frac{1}{2} \, \theta_I
\end{align}
(keep in mind, that the angle $\theta_I$ is negative). Again we
examine the OPE's of some special physical states with the fermionic
fields $\Psi(z)$ and $\bar{\Psi}(z)$. For $|\,0\rangle$ we get
\begin{align*}
\Psi^{I}(z)\, |\,0\rangle \, \rightarrow \,\, z^{\theta_I} \, t_I(0)
\qquad \qquad \bar{\Psi}^I(z)\,|\,0 \rangle \, \rightarrow \,\,
z^{-\theta_I} \, t_I'(0)
\end{align*}
and for $\psi_{-\frac{1}{2}-\theta_I}\, |\,0\rangle$
\begin{align*}
\Psi^I(z)\, \psi_{-\frac{1}{2}-\theta_I}\,|\,0\rangle \, \rightarrow
\,\,  z^{1+\theta_I} \, t_I(0)\qquad \qquad \bar{\Psi}^I(z)\,
\psi_{-\frac{1}{2}-\theta_I}\,|\,0\rangle \, \rightarrow
\,\,z^{-1-\theta_I}\, t_I'(0)\,\,.
\end{align*}
Before formulating the vertex operators for particular states we
also need the OPE's with the bosonic fields
\begin{align*}
\partial \, Z^I(z) \, |\,0\rangle  &\rightarrow z^{-(1-\theta_I)} \,\tau_I(0) \qquad
\qquad\,\,\, \bar{\partial} \, Z^I(\bar{z}) \, |\,0\rangle
\rightarrow z^{-\theta_I} \,
\tau_I(0)\\
\partial \, \bar{Z}^I(z) \, |\,0\rangle  &\rightarrow
z^{-\theta_I} \, \tau_I(0) \qquad \qquad \qquad \bar{\partial} \,
\bar{Z}^I(z) \, |\,0\rangle \rightarrow z^{-(1-\theta_I)} \,
\tau_I(0)\,\,.
\end{align*}
For negative angle, we replace $\theta_I$ by $\alpha_I=1+\theta_I$.\\
Now we can start to construct the vertex operators for the
respective states. First we consider the state
$\chi=\psi_{-\frac{1}{2}-\theta_3}\,|\,0\rangle$, where
$\theta_1$,$\theta_2$ are negative and $\theta_3$ is positive, which
means that the string starts at D-brane $a$ and ends at D-brane $b$
(see figure \ref{figure 1})\footnote{Recall that we count
counter-clockwise angles positive.}. The mass of this state is given
by
\begin{align*}
M^2=-\frac{1}{2} - \frac{1}{2} \, \theta_1 - \frac{1}{2} \, \theta_2
+ \frac{1}{2} \, \theta_3  + \frac{1}{2} - \theta_3 = -\sum^3_{I=1}
\theta_I\,\,.
\end{align*}
The scalar $\chi$ becomes massless when the sum of the angles adds
up to zero. This is in agreement with the supersymmetry condition.
The corresponding vertex operator in the (-1)-ghost picture takes
the form
\begin{align}
V_{\chi}^{-1}(z) = \mathrm{e}^{-\phi(z)}\,\prod^{2}_{I=1}
\sigma_{\theta_I}(z) \, \mathrm{e}^{\mathrm{i}\theta_I \,H_I (z)}\,
\sigma_{1+\theta_3}(z) \, \mathrm{e}^{\mathrm{i}(1+\theta_3)H_3
(z)}\, \mathrm{e}^{\mathrm{i} k \cdot X(z)} \,\, ,
\end{align}
where the $H_I$'s denote the bosonized worldsheet fermion $\Psi^I$.
Notice that in the case of supersymmetry, when the state becomes
massless ($k^2=0$), the conformal weight of the vertex
operator adds up, as required, to one. \\
The corresponding complex conjugate state $\chi^{*}$ is represented
by the same excitation as above but oriented from brane $b$ to brane
$a$. That means that the intersection angles $\theta'_I=-\theta_I$
take the opposite sign as before and therefore the vertex operator
is given by
\begin{align}
V_{\chi^{*}}^{-1}(z) = \mathrm{e}^{-\phi(z)}\,\prod^{2}_{I=1}
\sigma_{1-\theta_I}(z) \, \mathrm{e}^{-\mathrm{i}\theta_I \,H_I
(z)}\, \sigma_{-\theta_3}(z) \,
\mathrm{e}^{-\mathrm{i}(1+\theta_3)H_3 (z)}\, \mathrm{e}^{\mathrm{i}
k \cdot X(z)} \,\, ,
\end{align}
Let us take a closer look at the vertex operators in the case of
supersymmetry, when they carry a $N=2$ world sheet charge
$H=\sum^3_{I=1}H_I$. The chiral superfield $\chi$ has $N=2$ world
sheet charge +1, while the charge for the complex conjugate partner
$\chi^{*}$ is -1.\\
Next, we examine the state
$\chi^{*}=\psi_{-\frac{1}{2}+\theta_1}\,\psi_{-\frac{1}{2}+\theta_2}\,
\psi_{-\frac{1}{2}+\theta_3} |\,0 \rangle$, where $0<\theta_I<1$ for
all $I$. Again the string is oriented from brane $a$ to brane $b$
(see figure \ref{figure 2}). Why we denote the state by $\chi^{*}$
rather than $\chi$ becomes clear later. The mass of $\chi^{*}$ is
given by
\begin{align}
M^2=-\frac{1}{2}+\frac{1}{2}\,\sum^3_{I=1} \, \theta_I -\sum^3_{I=1}
\, \left(-\frac{1}{2}+\theta_I\right)=1-\frac{1}{2}\,\sum^3_{I=1} \,
\theta_I
\end{align}
and becomes massless, when the sum of the angles is equal to two,
again in agreement with the supersymmetry condition. The vertex
operator in the (-1)-ghost picture corresponding to this state takes
the form
\begin{align}
V^{(-1)}_{\chi^{*}}(z) = \mathrm{e}^{-\phi(z)}\,\prod^{3}_{I=1}
\sigma_{\theta_I}(z) \, \mathrm{e}^{\mathrm{i}(\theta_I-1)H_I (z)}\,
\mathrm{e}^{\mathrm{i} k \cdot X(z)}\,\, ,
\end{align}
and as above the requirement that the vertex operator has conformal
weight one is satisfied. The corresponding complex conjugated state
$\chi$ is stretched from brane $b$ to brane $a$ and the intersection
angles $\theta'_I=-\theta_I$ are all negative. Therefore the vertex
operator is given by
\begin{align}
V^{(-1)}_{\chi}(z) = \mathrm{e}^{-\phi(z)}\,\prod^{3}_{I=1}
\sigma_{1-\theta_I}(z) \, \mathrm{e}^{-\mathrm{i}(\theta_I-1)H_I
(z)}\, \mathrm{e}^{\mathrm{i} k \cdot X(z)}\,\, .
\end{align}
A look at the N=2 world sheet charge in the case of supersymmetry
($\sum^3_{I=1} \theta_I=2$) explains the notation since $\chi^{*}$
carries charge -1 while $\chi$ carries +1.\\
We now turn to the Ramond sector, in which the string excitations between two
intersecting D-branes correspond to space-time fermions. The mode expansion
for the fermionic degrees of freedom takes the same form as for the
Neveu-Schwarz (NS)- sector, but now we sum over integers instead of half
integers
\begin{align}
\Psi^I(z)= \sum_n \,\, \psi_{r-\theta_I}^I\,
z^{-r-\frac{1}{2}+\theta_I} \qquad \bar{\Psi}^I(z)= \sum_n \,\,
\psi_{r+\theta_I}^I\, \bar{z}^{-r-\frac{1}{2}-\theta_I}\,\,.
\end{align}
Nothing changes for the bosonic world sheet fields $Z(z,\bar{z})$
and $\bar{Z}(z,\bar{z})$. The vacuum is defined by ($0<\theta_I<1$)
\begin{equation}
\begin{aligned}
\alpha^I_{m-\theta_I} |\,0\rangle &=0   \qquad m \geq 1 \qquad
\qquad
\psi^I_{r-\theta_I} |\,0\rangle=0 \qquad r \geq 1 \\
\alpha^I_{m+\theta_I} |\,0\rangle &=0 \qquad m \geq 0\qquad \qquad
 \psi^I_{r+\theta_I} |\,0\rangle=0 \qquad r \geq 0\,\,.
\end{aligned}
\end{equation}
With this definition the zero point energy is independently of the
choice of angles given by
\begin{align}
\epsilon^I_0=0\,\,,
\end{align}
and therefore we always have a massless fermion in space time. While
the mass of the vacuum is independent on the angles the vertex
operator for the vacuum $|0\rangle$ depends crucially on the choice
of angles. Let us therefore examine the OPE's of worldsheet
fermions\footnote{The OPE with bosonic world-sheet fields is the
same as before for the NS-sector.} with $|0\rangle$ for the two
different situations that we have positive and negative intersecting
angles. We obtain for $0<\theta_I<1$
\begin{align*}
\Psi^I(z)\, |\,0\rangle \rightarrow z^{-\frac{1}{2} + \theta_I}
\,t_I(0)\qquad \qquad \bar{\Psi}^I(z)\, |\,0\rangle \rightarrow
z^{\frac{1}{2}- \theta_I} \,t_I(0)\,\,.
\end{align*}
For negative angles we must change the definition of the vacuum to
\begin{equation}
\begin{aligned}
\alpha^I_{m-\theta_I} |\,0\rangle &=0   \qquad m \geq 0\qquad \qquad
\psi^I_{r-\theta_I} |\,0\rangle=0 \qquad r \geq 0\\
\alpha^I_{m+\theta_I} |\,0\rangle &=0 \qquad m \geq 1\qquad \qquad
\psi^I_{r+\theta_I} |\,0\rangle=0 \qquad r \geq 1\,\,.
\end{aligned}
\end{equation}
The zero point energy is still zero. But now we obtain different
OPE's for the vacuum $|\,0\rangle$
\begin{align*}
\Psi^I(z)\, |\,0\rangle \rightarrow z^{\frac{1}{2} + \theta_I}
\,t_I(0) \qquad \qquad \bar{\Psi}^I(z)\, |\,0\rangle \rightarrow
z^{-\frac{1}{2}- \theta_I}\,t_I(0)\,\,.
\end{align*}
As before for the NS-sector we present for particular states the
vertex operators. The first state we consider is the vacuum state
$\chi=|\,0\rangle$, whose mass is independent of the choice of
angles equal to zero. Assuming, that the intersecting angles
$\theta_1$, $\theta_2$ in the first two internal dimensions are
positive and $\theta_3$ negative, the vertex operator takes the form
\begin{align}
V^{-\frac{1}{2}}_{\chi}(z)
 = \mathrm{e}^{-\frac{\phi}{2}(z)}\,S^{\alpha}(z) \prod^2_{I=1}
\sigma_{\theta_I} (z)\,
\mathrm{e}^{\mathrm{i}\left(\theta_I-\frac{1}{2}\right) H_I(z) } \,
\sigma_{1+\theta_3} (z)\,
\mathrm{e}^{\mathrm{i}\left(\theta_3+\frac{1}{2}\right) H_3(z) } \,
\mathrm{e}^{\mathrm{i} k \cdot X (z)} \,\, ,
\end{align}
where  $S^{\alpha}=\mathrm{e}^{\pm\textstyle{1\over 2}{\cal H}_1\pm
\textstyle{1\over 2}{\cal H}_2}$ denotes the spin field with
positive chirality\footnote{$\mathrm{e}^{{\cal H}_{1,2}}$ are the
bosonized world sheet fermions $\Psi^{a}$ where $a$ denotes the four
dimensional complexified indices.}. As for the NS-sector the
corresponding vertex operator for the complex conjugated state
$\chi^{*}$ is simply given by orientation reversal, so that the
intersection angles are $\theta'_I=-\theta_I$. Thus the vertex
operator in ($-\textstyle{1\over 2}$)-ghost picture has the form
\begin{align}
V^{-\frac{1}{2}}_{\chi^{*}}(z)
 = \mathrm{e}^{-\frac{\phi}{2}(z)}\,\tilde{S}_{\dot{\alpha}}(z) \prod^2_{I=1}
\sigma_{1-\theta_I} (z)\,
\mathrm{e}^{-\mathrm{i}\left(\theta_I-\frac{1}{2}\right) H_I(z) } \,
\sigma_{-\theta_3} (z)\,
\mathrm{e}^{-\mathrm{i}\left(\theta_3+\frac{1}{2}\right) H_3(z) } \,
\mathrm{e}^{\mathrm{i} k \cdot X (z)} \,\, ,
\end{align}
where $\tilde{S}_{\dot{\alpha}}=\mathrm{e}^{\pm\textstyle{1\over
2}{\cal H}_1\mp \textstyle{1\over 2}{\cal H}_2}$ represents the spin
field with opposite chirality as $S^{\alpha}$. Notice that
independent of the choice of angles the vertex operator has as
expected conformal weight one. As expected, in case of supersymmetry
($\sum^3_{I=3} \theta_I=0$) the vertex operators $\chi$ and
$\chi^{*}$ carry N=2 world sheet charge $-\textstyle{1\over 2}$ and
$\textstyle{1\over 2}$, respectively. \\
Finally let us assume that all the intersecting angles $\theta_I$
are positive. In that case the vertex operator for the vacuum state
$\chi^{*}$ takes a very symmetric form
\begin{align}
V_{\chi^{*}}^{-\frac{1}{2}}(z)
 = \mathrm{e}^{-\frac{\phi}{2}(z)}\,\tilde{S}_{\dot{\alpha}}(z)
\prod^3_{I=1} \sigma_{\theta_I} (z)\,
\mathrm{e}^{\mathrm{i}\left(\theta_I-\frac{1}{2}\right) H_I(z) } \,
\mathrm{e}^{\mathrm{i} k \cdot X (z)} \,\, .
\end{align}
For a similar reason as in the NS-sector we call this vacuum state
rather $\chi^{*}$ than $\chi$,since in case of supersymmetry
($\sum^3_{I=1}\theta_I=2$) it carries $\textstyle{1\over 2}$ N=2
world sheet charge. Following the procedure described above we
obtain for $\chi$
\begin{align}
V_{\chi}^{-\frac{1}{2}}(z)
 = \mathrm{e}^{-\frac{\phi}{2}(z)}\,S^{\alpha}(z)
\prod^3_{I=1} \sigma_{1-\theta_I} (z)\,
\mathrm{e}^{-\mathrm{i}\left(\theta_I-\frac{1}{2}\right) H_I(z) } \,
\mathrm{e}^{\mathrm{i} k \cdot X (z)} \,\,.
\end{align}
One can easily check that in case supersymmetry the vertex operator
carries as expected N=2 world sheet charge $H=-\frac{1}{2}$.

\newpage
\section{Numerical analysis}

Before we extract the low energy limit of the amplitudes, given
above, let us take a look at three different limits, namely $x
\rightarrow 0 $, $ x \rightarrow 1$ and $x\rightarrow -\infty$. The
first one  corresponds in the field theory to a gauge boson
exchange, while the latter one corresponds to a Higgs boson
exchange. In the limit $x\rightarrow 1$ the type of the exchange
particle  depends on which amplitude we examine; it is  either a
massive particle, for $\langle
{V^{\bar{5}}_{-\frac{1}{2}}}^{*}\,V^{\bar{5}}_{-\frac{1}{2}} \,
{V^{10}_{-\frac{1}{2}}}^{*} V^{10}_{-\frac{1}{2}} \rangle$ or again
a gauge boson for $\langle
{V^{10}_{-\frac{1}{2}}}^{*}\,V^{10}_{-\frac{1}{2}} \,
{V^{10}_{-\frac{1}{2}}}^{*} V^{10}_{-\frac{1}{2}} \rangle$. We start
with $\langle
{V^{\bar{5}}_{-\frac{1}{2}}}^{*}\,V^{\bar{5}}_{-\frac{1}{2}} \,
{V^{10}_{-\frac{1}{2}}}^{*} V^{10}_{-\frac{1}{2}} \rangle$ and turn
later to $\langle {V^{10}_{-\frac{1}{2}}}^{*}\,V^{10}_{-\frac{1}{2}}
\,
{V^{10}_{-\frac{1}{2}}}^{*} V^{10}_{-\frac{1}{2}} \rangle$.\\
\\
\emph{$x \rightarrow 0$}\\
\\
The limit $x \rightarrow 0$ was already explored in section 4 in
order to normalize the amplitude. Here we just state the result for
the case that $\theta_1=\theta_2=\theta$
\begin{itemize}
\item{$-\frac{1}{2}<\theta_{1}<0\qquad -\frac{1}{2}<\theta_{2}<0
\qquad \frac{1}{2}<\theta_{3}<1$\\
\\
\begin{align}
\sim \pi^{3/2} \int_0 \frac{\mathrm{dx}}{x}
\left[\left(\ln\left(\frac{\delta(\theta,1+2\theta)}{x}\right)
\right)^2
\ln\left(\frac{\delta(1+2\theta,-1-4\theta)}{x}\right)\right]^{-1}\,\,,
\label{asymptotic behavior}
\end{align}
where $\ln\delta(\theta,\nu)$ is given by
\begin{align}
\ln\delta(\theta,\nu)=2\psi(1)-\frac{1}{2}\psi(\theta)-
\frac{1}{2}\psi(1-\theta)-\frac{1}{2}\psi(\nu)-\frac{1}{2}\psi(1-\nu)\,\,.
\label{delta}
\end{align}
} \item{$\frac{1}{2}<\theta_{1}<1\qquad \frac{1}{2}<\theta_{2}<1
\qquad \frac{1}{2}<\theta_{3}<1$
\\
\\
\begin{align}
\sim \pi^{3/2} \int_0 \frac{\mathrm{dx}}{x}
\left[\left(\ln\left(\frac{\delta(-\theta,-1+2\theta)}{x}\right)
\right)^2
\ln\left(\frac{\delta(-1+2\theta,3-4\theta)}{x}\right)\right]^{-1}
\end{align}
with the same $\delta(\theta,\nu)$ as above.  }
\\
\\

\end{itemize}
\emph{$x \rightarrow 1$}\\
\\
Using the properties of the Hypergeometric function, in particular
the transformation law
\begin{align*}\nonumber
_2F_1(a,b,c;\, x)&=\frac{\Gamma(c)\,\Gamma(c-a-b)}{\Gamma(c-a)\,\Gamma(c-b)}\,\, _2F_1(a,b,a+b-c+1;1-x)\\
\\ \nonumber
&+(1-z)^{c-a-b}\,
\frac{\Gamma(c)\,\Gamma(a+b-c)}{\Gamma(a)\,\Gamma(b)}\,\,
_2F_1(c-a,c-b,c-a-b-1;1-x)
\end{align*}
and the limit
\begin{align*}
\lim_{x\rightarrow0} \, _2F_1(a,b,c;\,x) = 1\,\,,
\end{align*}
we obtain
\begin{align}
\lim_{x \rightarrow 0} \frac{1}{2\pi}\, I^{-1}(a,b,x)
\longrightarrow
\begin{cases}
\frac{\Gamma(1-a)\,\Gamma(b)\,\Gamma(1+a-b)}{\Gamma(a)\,\Gamma(1-b)\,\Gamma(b-a)}\,\,
(1-x)^{b-a}
 & \hspace{0.1cm}
a<b  \\
\frac{\Gamma(a)\,\Gamma(1-b)\,\Gamma(1-a+b)}{\Gamma(1-a)\,\Gamma(b)\,\Gamma(a-b)}\,\,
(1-x)^{a-b} & \hspace{0.1cm} a>b
\end{cases}\,\,.
\end{align}
In this limit we do not obtain an integer mode, which tells us that
the exchange particle is massive. The mass depends on the choice of
angles, as we will show based on our first case
($-\frac{1}{2}<\theta_{1}<0\,, -\frac{1}{2}<\theta_{2}<0 \,,
\frac{1}{2}<\theta_{3}<1$). Let us assume that the two angles
$\theta_1$ and $\theta_2$ are equal
\begin{align}
\theta_1=\theta_2=\theta \rightarrow \theta_3=-2\theta\,\,.
\end{align}
In the limit $x\rightarrow 1$ the amplitude \eqref{final result
551010 1} takes the form
\begin{align}
\sim \frac{\Gamma(1+\theta)\Gamma(1+2\theta)\Gamma(-3\theta)}
{\Gamma(-\theta)\Gamma(-2\theta)\Gamma(1+3\theta)} \sqrt{
\frac{\Gamma(1+2\theta)\Gamma(2+4\theta)\Gamma(-1-6\theta)}
{\Gamma(-2\theta)\Gamma(-1-4\theta)\Gamma(2+6\theta)}} \, \int^{1}
(1-x)^{-\alpha'u+1+6\theta} \label{massive Higgs1}
\end{align}
for $\theta>-1/3$ and
\begin{align}
\sim \frac{\Gamma(-\theta)\Gamma(-2\theta)\Gamma(2+3\theta)}
{\Gamma(1+\theta)\Gamma(1+2\theta)\Gamma(-1-3\theta)}
 \sqrt{
\frac{\Gamma(-2\theta)\Gamma(-1-4\theta)\Gamma(3+6\theta)}
{\Gamma(1+2\theta)\Gamma(2+4\theta)\Gamma(-2-6\theta)} } \, \int^{1}
(1-x)^{-\alpha'u-3-6\theta}
 \label{massive Higgs2}
\end{align}
for $\theta<-1/3$. In the low energy limit \eqref{massive Higgs1}
and \eqref{massive Higgs2} are proportional to
\begin{align}
\emph{A}\sim\frac{1}{\alpha'u-\alpha'M^2}\,\,,
\end{align}
where $M$ denotes the mass of the exchanged particle and is given by
\begin{align}
\alpha'M^2=
\begin{cases}
2+6\theta
 & \hspace{0.1cm}
\theta>-\frac{1}{3}  \\
-2-6\theta & \hspace{0.1cm} \theta<-\frac{1}{3}
\end{cases},
\end{align}
which becomes massless for $\theta=1/3$. For this choice of angle we
observe $N=2$ supersymmetry in the Minkowski-space. Since we focus
on models with  N=1 chiral fermion sector, only,  we do not take
this limit.   For our second amplitude \eqref{final result 551010 2}
we also observe a massive particle exchange in this limit
\begin{align}
\sim \frac{\Gamma(\theta)\Gamma(-1+2\theta)\Gamma(3-3\theta)}
{\Gamma(1-\theta)\Gamma(2-2\theta)\Gamma(-2+3\theta)} \sqrt{
\frac{\Gamma(-1+2\theta)\Gamma(-2+4\theta)\Gamma(5-6\theta)}
{\Gamma(2-2\theta)\Gamma(3-4\theta)\Gamma(-4+6\theta)}} \, \int^{1}
(1-x)^{-\alpha'u-5+6\theta}
\end{align}
for $\theta>2/3$ and
\begin{align}
\sim \frac{\Gamma(1-\theta)\Gamma(2-2\theta)\Gamma(-1+3\theta)}
{\Gamma(\theta)\Gamma(-1+2\theta)\Gamma(2-3\theta)}
 \sqrt{
\frac{\Gamma(2-2\theta)\Gamma(3-4\theta)\Gamma(-3+6\theta)}
{\Gamma(-1+2\theta)\Gamma(-2+4\theta)\Gamma(4-6\theta)} } \,
\int^{1} (1-x)^{-\alpha'u+3-6\theta}
\end{align}
for $\theta<2/3$. In our effective low energy theory we integrate
out all massive states, so that the part of the amplitude arising
from these string massive state exchanges contribute to the four-Fermi contact term.\\
\\
\emph{$x\rightarrow -\infty$}\\
\\
At last let us examine the limit $x \rightarrow -\infty$. As
mentioned earlier the second terms of \eqref{final result 551010 1}
and \eqref{final result 551010 2} give the contribution to the four
fermi interaction arising from the massless Higgs particle exchange.
Therefore in the limit $x\rightarrow -\infty$ we expect
to observe an exchange of a massless particle.\\
The hypergeometric functions behave in the limit $ x\rightarrow
-\infty $
\begin{align*}
\nonumber \lim_{x \rightarrow -\infty} F(a,b,c,x)&=
\frac{\Gamma(c)\,\Gamma(b-a)}{\Gamma(b)\,\Gamma(c-a)}\, x^{-a} +
\frac{\Gamma(c)\,\Gamma(a-b)}{\Gamma(a)\,\Gamma(c-b)}\, x^{-b}\\
\\
\nonumber \lim_{x \rightarrow -\infty} F(a,b,c,1-x)  &=
\mathrm{e}^{-\mathrm{i} \pi a}\,\,
\frac{\Gamma(c)\,\Gamma(b-a)}{\Gamma(b)\,\Gamma(c-a)}\, x^{-a} +
\mathrm{e}^{-\mathrm{i} \pi b}\,\,
\frac{\Gamma(c)\,\Gamma(a-b)}{\Gamma(a)\,\Gamma(c-b)}\, x^{-b} \,\,.
\end{align*}
Hence $I(a,b,x)$ for $x \rightarrow -\infty$ takes the form
\begin{align}
\lim_{x \rightarrow -\infty } \frac{1}{2\pi}I_j(a,b,x)^{-1}
\longrightarrow
\begin{cases}
(-1)^{a -b}\,  x^{a+b} \, \Gamma_{a,b} & \hspace{0.1cm}
0<a+b <1 \\
-(-1)^{a-b}\,   x^{2-a-b}\, \Gamma_{1-a,1-b} & \hspace{0.1cm} 1<a+b
<2
\end{cases},
\label{limit}
\end{align}
with
\begin{align}
\Gamma_{a,b} = \frac{\Gamma(1-a)\, \Gamma(1-b)\,\Gamma(a+
b)}{\Gamma(a)\, \Gamma(b)\,\Gamma(1-a-b) }\,\,.
\end{align}
Using \eqref{limit} the amplitude \eqref{final result 551010 1}
becomes in the limit $x\rightarrow -\infty$
\begin{align}
\sim (2\pi)^{\frac{3}{2}}\,
\Gamma^{\frac{1}{2}}_{\theta_1,1-2\theta_1}\,\Gamma^{\frac{1}{2}}_{\theta_2,1-2\theta_2}
 \, \Gamma^{\frac{1}{2}}_{1+\theta_3,-1-2\theta_3}\,\int_{-\infty}\,
\mathrm{d}x \,\,x^{-\alpha't-1} \,\,.\label{Higgs exchange}
\end{align}
Thus, we observe an exchange of a massless particle, which we
identify as the Higgs-particle. Note that the prefactor in
\eqref{Higgs exchange} is the expected relative factor between the
Yukawa couplings in string and field theory basis 
\cite{{Lust:2004cx},{Cvetic:2003ch},{Bertolini:2005qh}} .\\
Applying the limit for our second amplitude \eqref{final result
551010 2} we obtain
\begin{align}
\sim (2\pi)^{\frac{3}{2}}\, \prod^3_{I=1}
\Gamma^{\frac{1}{2}}_{1+\theta_I,-1-2\theta_I} \,\int_{-\infty}\,
\mathrm{d}x \,\,x^{-\alpha't-1}
\end{align}
and again we can observe a massless Higgs exchange in this
limit.\\
\\
\textbf{The amplitude $\langle
{V^{\bar{5}}_{-\frac{1}{2}}}^{*}\,V^{\bar{5}}_{-\frac{1}{2}} \,
{V^{10}_{-\frac{1}{2}}}^{*} V^{10}_{-\frac{1}{2}} \rangle$}\\
\\
The analysis for both amplitudes, \eqref{final result 551010 1} and
\eqref{final result 551010 2} is similar, so that we will describe
the steps for the first one and apply these later for the second
amplitude. We start by investigating the integral
$K(\theta_1,\theta_2,\theta_3)$ and turn later to
$T(\theta_1,\theta_2,\theta_3)$.\\
\\
\emph{$K(\theta_1,\theta_2,\theta_3)$}\\
\\
Since in this interval the amplitude is finite even in the low
energy limit, we do not have to subtract anything. Thus, we can send
$\alpha'$ to zero and obtain
\begin{align}
K=\int^1_0 \mathrm{d}x \frac{1}{x(1-x)} \left[I\left(-\theta_1,
1+\nu_1, x\right) \,I\left(-\theta_2, 1+\nu_2, x\right)
\,I\left(1-\theta_3, 1+\nu_3, x\right)\right]^{-\frac{1}{2}}\,\,.
\label{K}
\end{align}
Let us split the integral \eqref{K} by using the expression
\begin{align}
\frac{1}{x(1-x)} = \frac{1}{x} + \frac{1}{1-x}\,\,.
\label{splitting}
\end{align}
Let us first evaluate the integral starting with the first summand
of \eqref{K} which is given by
\begin{align}
K_1=\int^1_0 \mathrm{d}x \frac{1}{x} \left[I\left(-\theta_1,
1+\nu_1, x\right) \,I\left(-\theta_2, 1+\nu_2, x\right)
\,I\left(1-\theta_3, 1+\nu_3, x\right)\right]^{-\frac{1}{2}}\,\,.
\end{align}
Substituting $\mathrm{e}^{-t}$ for $x$ we obtain
\begin{align}
K_1=\int^{\infty}_0 \mathrm{d}t
 \left[I\left(-\theta_1, 1+\nu_1,
\mathrm{e}^{-t}\right) \,I\left(-\theta_2, 1+\nu_2,
\mathrm{e}^{-t}\right) \,I\left(1-\theta_3, 1+\nu_3,
\mathrm{e}^{-t}\right)\right]^{-\frac{1}{2}}\,\,. \label{K(theta)}
\end{align}
Mathematica is not able to evaluate this expression numerically
since it is hard to maintain numerical precision for large $t$.
Therefore we will split integral \eqref{K(theta)} into the range
from $0$ to $T$ and from $T$ to $\infty$. For the computation of the
first region we will use Mathematica to evaluate it numerically,
while for the second region we replace the hypergeometric functions
by their asymptotic behavior given in \eqref{asymptotic behavior}
\begin{align*}
K_1 &=\int^{T}_0 \mathrm{d}t
 \left[I\left(-\theta_1, 1+\nu_1,
\mathrm{e}^{-t}\right) \,I\left(-\theta_2, 1+\nu_2,
\mathrm{e}^{-t}\right) \,I\left(1-\theta_3, 1+\nu_3,
\mathrm{e}^{-t}\right)\right]^{-\frac{1}{2}} \\
&+ \pi^{3/2}\,\int^{\infty}_T \mathrm{d} t
\left[\left(t+\ln\delta(-\theta_1,1+\nu_1)\right)
\left(t+\ln\delta(-\theta_2,1+\nu_2)\right)
\left(t+\ln\delta(1-\theta_3,1+\nu_3)\right)\right]^{-\frac{1}{2}}\,\,.
\end{align*}
Let us assume that the two angles $\theta_1$ and $\theta_2$ are
equal to each other
\begin{align*}
\theta_1=\theta_2=\theta \rightarrow \theta_3=-2\theta\,\,.
\end{align*}
Then $K_1$ simplifies to
\begin{align}
K_1=&\int^{T}_0 \mathrm{d}t
 \left[I\left(-\theta, 1+2\theta,
\mathrm{e}^{-t}\right) \,I\left(-\theta, 1+2\theta,
\mathrm{e}^{-t}\right) \,I\left(1+2\theta, -1-4\theta,
\mathrm{e}^{-t}\right)\right]^{-\frac{1}{2}}
\\ \nonumber
& \qquad+ \pi^{3/2}\,\int^{\infty}_T \mathrm{d}
t\left(t+\ln\delta(-\theta,1+2\theta)\right)^{-1}
\left(t+\ln\delta(1+2\theta,-1-4\theta)\right)^{-\frac{1}{2}}\,\,.
\end{align}
Now we turn to the second term we get after splitting the integral.
Again we substitute $\mathrm{e}^{-t}$ for $x$, set, as above,
$\theta_1=\theta_2=\theta$ and obtain
\begin{align*}
K_2=\int^{\infty}_0 \mathrm{d}t \,
\frac{\mathrm{e}^{-t}}{1-\mathrm{e}^{-t}}\,
 \left[I^2\left(-\theta, 1+2\theta,\mathrm{e}^{-t}\right)  \,I\left(1+2\theta, -1-4\theta,
\mathrm{e}^{-t}\right)\right]^{-\frac{1}{2}}\,\,.
\end{align*}
As above we have to split this integral into two parts, where we
replace the $I$'s by their asymptotic behavior
\begin{align}  K_2&=\int^{T}_0 \mathrm{d}t \,
\frac{\mathrm{e}^{-t}}{1-\mathrm{e}^{-t}}\,
 \left[I^2\left(-\theta, 1+2\theta,\mathrm{e}^{-t}\right)  \,I\left(1+2\theta, -1-4\theta,
\mathrm{e}^{-t}\right)\right]^{-\frac{1}{2}} \label{general
numerical expression2}
\\ \nonumber
& \qquad+ \pi^{3/2}\,\int^{\infty}_T \mathrm{d} t\,
\frac{\mathrm{e}^{-t}}{1-\mathrm{e}^{-t}}\,
\left(t+\ln\delta(-\theta,1+2\theta)\right)^{-1}
\left(t+\ln\delta(1+2\theta,-1-4\theta)\right)^{-\frac{1}{2}}\,\,.
\end{align}
Applying the same procedure for the other sector we obtain
\begin{itemize}
\item{$\frac{1}{2}<\theta_{1}<1\qquad \frac{1}{2}<\theta_{2}<1
\qquad \frac{1}{2}<\theta_{3}<1$ \\
\\
In this sector we obtain
\begin{align}
K_1=&\int^{T}_0 \mathrm{d}t
 \left[I^{2}\left(1-\theta, -1+2\theta,
\mathrm{e}^{-t}\right) \,I\left(2\theta-1, 3-4\theta,
\mathrm{e}^{-t}\right)\right]^{-\frac{1}{2}} \label{general
numerical expression3}
\\ \nonumber
& \qquad+ \pi^{3/2}\,\int^{\infty}_T \mathrm{d}
t\left(t+\ln\delta(1-\theta,-1+2\theta)\right)^{-1}
\left(t+\ln\delta(2\theta-1,3-4\theta)\right)^{-\frac{1}{2}}
\end{align}
and
\begin{align}  K_2&=\int^{T}_0 \mathrm{d}t \,
\frac{\mathrm{e}^{-t}}{1-\mathrm{e}^{-t}}\,
 \left[I^{2}\left(1-\theta, -1+2\theta,\mathrm{e}^{-t}\right)  \,I\left(2\theta-1, 3-4\theta,
\mathrm{e}^{-t}\right)\right]^{-\frac{1}{2}} \label{general
numerical expression2}
\\ \nonumber
& \qquad + \pi^{3/2}\,\int^{\infty}_T \mathrm{d} t\,
\frac{\mathrm{e}^{-t}}{1-\mathrm{e}^{-t}}\,
\left(t+\ln\delta(1-\theta,-1+2\theta)\right)^{-1}
\left(t+\ln\delta(2\theta-1,3-4\theta)\right)^{-\frac{1}{2}}\,\,.
\end{align}
}
\end{itemize}
The whole integral $K(\theta)$ is given by the sum of $K_1$ and
$K_2$.\\
\\
\emph{ $T(\theta_1,\theta_2,\theta_3)$}\\
\\
Let us now analyze the massive string state contribution to
$T(\theta_1,\theta_2,\theta_3)$, where  in the field theory the
proton decay takes place via Higgs particle mediation. Thus, in
contrast to the numerical analysis for proton decay via a gauge
boson exchange we observe a pole that corresponds to the Higgs
exchange. In order to obtain the four-Fermi interaction term due to
the massive string states, we need to
subtract this pole before taking the low energy limit.\\
Let us split the integral \eqref{definition of K and T} into two
parts (again we assume that $\theta_1=\theta_2=\theta$)
\begin{align}
 \int^{L}_{-\infty} \, \mathrm{d}x \,x^{-\alpha's-1}\,
(1-x)^{-\alpha'u-1} \, \left[I^2\left(-\theta,1+2\theta,x\right)\,
I\left(1+2\theta,-1-4\theta,x \right)\right]^{-\frac{1}{2}}\\
\nonumber+ \int^{0}_{L} \, \mathrm{d}x \,x^{-\alpha's-1}\,
(1-x)^{-\alpha'u-1} \, \left[I^2\left(-\theta,1+2\theta,x\right)\,
I\left(1+2\theta,-1-4\theta,x \right)\right]^{-\frac{1}{2}}\,\,\, .
\end{align}
Now we replace $x$ by $1-e^{z}$ in the first summand and in the
second by $\frac{1}{1-e^{z}}$
\begin{align}
 \int^{\infty}_{\ln(1-L)} \, \mathrm{d}z
 \frac{ \left[I^2\left(-\theta,1+2\theta,1-e^z\right)\,
I\left(1+2\theta,-1-4\theta,1-e^z
\right)\right]^{-\frac{1}{2}}}{(e^z)^{\alpha'u}\,
(1-e^z)^{\alpha's+1}}
\\ \nonumber
+ \int^{\infty}_{\ln(1-\frac{1}{L})} \, \mathrm{d}z \frac{
\left[I^2\left(-\theta,1+2\theta,\frac{1}{1-e^z}\right)\,
I\left(1+2\theta,-1-4\theta,\frac{1}{1-e^z}
\right)\right]^{-\frac{1}{2}}}{ (e^z)^{\alpha'u}\,
(1-e^z)^{\alpha't} }\,\,\, .
\end{align}
To simplify the computation, we break up both terms into two parts
\begin{align*}
 &\int^{T_1}_{\ln(1-L)} \, \mathrm{d}z
 \frac{ \left[I^2\left(-\theta,1+2\theta,1-e^z\right)\,
I\left(1+2\theta,-1-4\theta,1-e^z
\right)\right]^{-\frac{1}{2}}}{(e^z)^{\alpha'u}\,
(1-e^z)^{\alpha's+1}}
\\ \nonumber
 + &\,\,(2 \pi)^{\frac{3}{2}} \, \Gamma_{-\theta,1+2\theta} \,
\Gamma^{\frac{1}{2}}_{1+2\theta,-1-4\theta} \int^{\infty}_{T1}
\mathrm{d} z \, (e^{z})^{-\alpha'u+1} \, (1-e^{z})^{-\alpha's-1}
\\
+ &\int^{T_2}_{\ln(1-\frac{1}{L})} \, \mathrm{d}z \, \frac{
\left[I^2\left(-\theta,1+2\theta,\frac{1}{1-e^z}\right)\,
I\left(1+2\theta,-1-4\theta,\frac{1}{1-e^z}
\right)\right]^{-\frac{1}{2}}} {  (e^z)^{\alpha'u}\, (1-e^z)^{\alpha't} }\\
+&\,\,\pi^{\frac{3}{2}} \int^{\infty}_{T_2} \mathrm{d}z \frac{
\,\left(z+\ln\delta(-\theta,1+2\theta)\right)^{-1} \left(
z+\ln\delta(1+2\theta,-1-4\theta)\right)^{-\frac{1}{2}}}{
(e^z)^{\alpha'u}\, (1-e^z)^{\alpha't} }
\end{align*}
Here we replaced the hypergeometric expressions by their respective
limits in the range from $T_1$ to $\infty$ and $T_2$ to $\infty$ .
As mentioned above in order to get the four-Fermi interaction
contribution, we need to subtract the $\frac{1}{\alpha't}$ pole and
take the low energy limit
\begin{align*}
T(\theta) = \lim_{\alpha'\rightarrow 0 }\Big\{
 &\int^{T_1}_{\ln(1-L)} \, \mathrm{d}z\,
\frac{ \left[I^2\left(-\theta,1+2\theta,1-e^z\right)\,
I\left(1+2\theta,-1-4\theta,1-e^z
\right)\right]^{-\frac{1}{2}}}{(e^z)^{\alpha'u}\,
(1-e^z)^{\alpha's+1}}
\\ \nonumber
&  + \left((2 \pi)^{\frac{3}{2}} \, \Gamma_{-\theta,1+2\theta} \,
\Gamma^{\frac{1}{2}}_{1+2\theta,-1-4\theta} \int^{\infty}_{T1}
\mathrm{d} z \, (e^{z})^{-\alpha'u+1} \, (1-e^{z})^{-\alpha's-1}
-\frac{1}{\alpha't} \right)
\\
& + \int^{T_2}_{\ln(1-\frac{1}{L})} \, \mathrm{d}z \, \frac{
\left[I^2\left(-\theta,1+2\theta,\frac{1}{1-e^z}\right)\,
I\left(1+2\theta,-1-4\theta,\frac{1}{1-e^z}
\right)\right]^{-\frac{1}{2}}}{(e^z)^{\alpha'u}\, (1-e^z)^{\alpha't}  }\\
&+\pi^{\frac{3}{2}} \int^{\infty}_{T_2} \mathrm{d}z\, \frac{
\left(z+\ln\delta(-\theta,1+2\theta)\right)^{-1} \left(
z+\ln\delta(1+2\theta,-1-4\theta)\right)^{-\frac{1}{2}}}{(e^z)^{\alpha'u}\,
(1-e^z)^{\alpha't}} \Big\}\,\,.
\end{align*}
For the second region $\frac{1}{2}<\theta_{1}<1$,
$\frac{1}{2}<\theta_{2}<1$ and $\frac{1}{2}<\theta_{3}<1$,
$T(\theta)$ takes the form
\begin{align*}
T(\theta) = \lim_{\alpha'\rightarrow 0 }\Big\{
 &\int^{T_1}_{\ln(1-L)} \, \mathrm{d}z\,
\frac{ \left[I^2\left(1-\theta,-1+2\theta,1-e^z\right)\,
I\left(-1+2\theta,3-4\theta,1-e^z
\right)\right]^{-\frac{1}{2}}}{(e^z)^{\alpha'u}\,
(1-e^z)^{\alpha's+1}}
\\ \nonumber
&  + \left((2 \pi)^{\frac{3}{2}} \, \Gamma_{1-\theta,-1+2\theta} \,
\Gamma^{\frac{1}{2}}_{-1+2\theta,3-4\theta} \int^{\infty}_{T1}
\mathrm{d} z \, (e^{z})^{-\alpha'u+1} \, (1-e^{z})^{-\alpha's-1}
-\frac{1}{\alpha't} \right)
\\
& + \int^{T_2}_{\ln(1-\frac{1}{L})} \, \mathrm{d}z \, \frac{
\left[I^2\left(1-\theta,-1+2\theta,\frac{1}{1-e^z}\right)\,
I\left(-1+2\theta,3-4\theta,\frac{1}{1-e^z}
\right)\right]^{-\frac{1}{2}}}{(e^z)^{\alpha'u}\, (1-e^z)^{\alpha't}  }\\
&+\pi^{\frac{3}{2}} \int^{\infty}_{T_2} \mathrm{d}z\, \frac{
\left(z+\ln\delta(1-\theta,-1+2\theta)\right)^{-1} \left(
z+\ln\delta(-1+2\theta,3-4\theta)\right)^{-\frac{1}{2}}}{(e^z)^{\alpha'u}\,
(1-e^z)^{\alpha't}} \Big\}\,\,.
\end{align*}
Mathematica is not able to take that limit, however by plugging in
different small values for $\alpha'$ (keep in mind that the
Mandelstam variables $s$, $t$ and
$u$ have to satisfy momentum conservation $s+t+u=0$) we get a stable contribution for $T(\theta)$.\\
\\
\textbf{The amplitude $\langle
{V^{10}_{-\frac{1}{2}}}^{*}\,V^{10}_{-\frac{1}{2}} \,
{V^{10}_{-\frac{1}{2}}}^{*} V^{10}_{-\frac{1}{2}} \rangle$}\\
\\
The analysis is simpler for $\langle
{V^{\bar{5}}_{-\frac{1}{2}}}^{*}\,V^{\bar{5}}_{-\frac{1}{2}} \,
{V^{10}_{-\frac{1}{2}}}^{*} V^{10}_{-\frac{1}{2}} \rangle$ because
of the symmetry of the amplitude:  after splitting the integral
\eqref{splitting} both parts give the same contribution, so that we
only need to focus on one part and multiply by a factor of two.
Following the same steps as above the integral $M$ becomes
\begin{align*}
 M&=2\, \int^{T}_0 \mathrm{d}t
\prod^3_{I=1} \sqrt{\sin[\pi(1+\nu_I)]} \,\,
L^{-\frac{1}{2}}(1+\nu_I)
+ \pi^{3/2}\,\int^{\infty}_T \mathrm{d}t\, \prod^3_{I=1}
\left(t+\ln\delta(1+\nu_I,1+\nu_I)\right)^{-\frac{1}{2}}
\end{align*}
Replacing $\nu_I$ by $\theta_I$ and assuming that
$\theta_1=\theta_2$ we get for
\begin{itemize}
\item{$-\frac{1}{2}<\theta_{1}<0\qquad -\frac{1}{2}<\theta_{2}<0
\qquad \frac{1}{2}<\theta_{3}<1$\\
\begin{align}
 M&=2\, \int^{T}_0 \mathrm{d}t\,
 \sin[\pi(1+2\theta)] \sqrt{\sin[\pi(-1-4\theta)]} \,\,
L^{-1}(1+2\theta) \,  L^{-\frac{1}{2}}(-1-4\theta)  \\ \nonumber &
\qquad \qquad  + \pi^{3/2}\,\int^{\infty}_T \mathrm{d}t\,
\left(t+\ln\delta(1+2\theta,1+2\theta)\right)^{-1}
\left(t+\ln\delta(-1-4\theta,-1-4\theta)\right)^{-\frac{1}{2}}\,\,,
\end{align}
and for  } \item{$\frac{1}{2}<\theta_{1}<1\qquad
\frac{1}{2}<\theta_{2}<1 \qquad \frac{1}{2}<\theta_{3}<1$\\
\begin{align}
 M&=2\, \int^{T}_0 \mathrm{d}t\,
 \sin[\pi(2\theta-1)] \sqrt{\sin[\pi(3-4\theta)]} \,\,
L^{-1}(2\theta-1) \,  L^{-\frac{1}{2}}(3-4\theta)  \\ \nonumber &
\qquad \qquad  + \pi^{3/2}\,\int^{\infty}_T \mathrm{d}t\,
\left(t+\ln\delta(2\theta-1,2\theta-1)\right)^{-1}
\left(t+\ln\delta(3-4\theta,3-4\theta)\right)^{-\frac{1}{2}}\,\,.
\end{align}

}
\end{itemize}

\newpage

\bibliographystyle{JHEP}
\bibliography{proton}

\providecommand{\href}[2]{#2}\begingroup\raggedright\begin{thebibliography}{10}

\bibitem{Georgi:1974sy}
H.~Georgi and S.~L. Glashow, {\it Unity of all elementary particle forces},
  {\em Phys. Rev. Lett.} {\bf 32} (1974) 438--441.

\bibitem{Langacker:1980js}
P.~Langacker, {\it Grand unified theories and proton decay},  {\em Phys. Rept.}
  {\bf 72} (1981) 185.

\bibitem{Nath:2006ut}
P.~Nath and P.~F. Per{\'e}z, {\it Proton stability in grand unified theories,
  in strings, and in branes},
  \href{http://xxx.lanl.gov/abs/hep-ph/0601023}{{\tt hep-ph/0601023}}.

\bibitem{Sakai:1981gr}
N.~Sakai, {\it Naturalness in supersymmetric {'GUTS'}},  {\em Zeit. Phys.} {\bf
  C11} (1981) 153.

\bibitem{Dimopoulos:1981zb}
S.~Dimopoulos and H.~Georgi, {\it Softly broken supersymmetry and {SU(5)}},
  {\em Nucl. Phys.} {\bf B193} (1981) 150.

\bibitem{Murayama:2001ur}
H.~Murayama and A.~Pierce, {\it Not even decoupling can save minimal
  supersymmetric {SU(5)}},  {\em Phys. Rev.} {\bf D65} (2002) 055009,
  [\href{http://xxx.lanl.gov/abs/hep-ph/0108104}{{\tt hep-ph/0108104}}].

\bibitem{Dermisek:2000hr}
R.~Derm{\'i}s{\v{e}}k, A.~Mafi, and S.~Raby, {\it {SUSY GUTs} under siege:
  Proton decay},  {\em Phys. Rev.} {\bf D63} (2001) 035001,
  [\href{http://xxx.lanl.gov/abs/hep-ph/0007213}{{\tt hep-ph/0007213}}].

\bibitem{Hisano:2000dg}
J.~Hisano, {\it Proton decay in the supersymmetric grand unified models},
  \href{http://xxx.lanl.gov/abs/hep-ph/0004266}{{\tt hep-ph/0004266}}.

\bibitem{Emmanuel-Costa:2003pu}
D.~Emmanuel-Costa and S.~Wiesenfeldt, {\it Proton decay in a consistent
  supersymmetric {SU(5) GUT} model},  {\em Nucl. Phys.} {\bf B661} (2003)
  62--82, [\href{http://xxx.lanl.gov/abs/hep-ph/0302272}{{\tt
  hep-ph/0302272}}].

\bibitem{Bajc:2002pg}
B.~Bajc, P.~Fileviez~Per{\'e}z, and G.~Senjanov{\'i}c, {\it Minimal
  supersymmetric {SU(5)} theory and proton decay: Where do we stand?},
  \href{http://xxx.lanl.gov/abs/hep-ph/0210374}{{\tt hep-ph/0210374}}.

\bibitem{Bajc:2002bv}
B.~Bajc, P.~Fileviez~Per{\'e}z, and G.~Senjanov{\'i}c, {\it Proton decay in
  minimal supersymmetric {SU(5)}},  {\em Phys. Rev.} {\bf D66} (2002) 075005,
  [\href{http://xxx.lanl.gov/abs/hep-ph/0204311}{{\tt hep-ph/0204311}}].

\bibitem{FileviezPerez:2004hn}
P.~Fileviez~Per{\'e}z, {\it Fermion mixings vs d = 6 proton decay},  {\em Phys.
  Lett.} {\bf B595} (2004) 476--483,
  [\href{http://xxx.lanl.gov/abs/hep-ph/0403286}{{\tt hep-ph/0403286}}].

\bibitem{Cvetic:2002pj}
M.~Cveti{\v{c}}, I.~Papadimitriou, and G.~Shiu, {\it Supersymmetric three
  family {SU(5)} grand unified models from type {IIA} orientifolds with
  intersecting {D}6-branes},  {\em Nucl. Phys.} {\bf B659} (2003) 193--223,
  [\href{http://xxx.lanl.gov/abs/hep-th/0212177}{{\tt hep-th/0212177}}].

\bibitem{Cvetic:2001nr}
M.~Cveti{\v{c}}, G.~Shiu, and A.~M. Uranga, {\it Chiral four-dimensional {N =
  1} supersymmetric type {IIA} orientifolds from intersecting {D6}-branes},
  {\em Nucl. Phys.} {\bf B615} (2001) 3--32,
  [\href{http://xxx.lanl.gov/abs/hep-th/0107166}{{\tt hep-th/0107166}}].

\bibitem{Cvetic:2001tj}
M.~Cveti{\v{c}}, G.~Shiu, and A.~M. Uranga, {\it Three-family supersymmetric
  standard like models from intersecting brane worlds},  {\em Phys. Rev. Lett.}
  {\bf 87} (2001) 201801, [\href{http://xxx.lanl.gov/abs/hep-th/0107143}{{\tt
  hep-th/0107143}}].

\bibitem{Blumenhagen:2005mu}
R.~Blumenhagen, M.~Cveti{\v{c}}, P.~Langacker, and G.~Shiu, {\it Toward
  realistic intersecting {D}-brane models},
  \href{http://xxx.lanl.gov/abs/hep-th/0502005}{{\tt hep-th/0502005}}.

\bibitem{Aldazabal:2000cn}
G.~Aldazabal, S.~Franco, L.~E. Ib{\'a}{\~n}ez, R.~Rabad{\'a}n, and A.~M.
  Uranga, {\it Intersecting brane worlds},  {\em JHEP} {\bf 02} (2001) 047,
  [\href{http://xxx.lanl.gov/abs/hep-ph/0011132}{{\tt hep-ph/0011132}}].

\bibitem{Blumenhagen:2001te}
R.~Blumenhagen, B.~K{\"o}rs, D.~L{\"u}st, and T.~Ott, {\it The standard model
  from stable intersecting brane world orbifolds},  {\em Nucl. Phys.} {\bf
  B616} (2001) 3--33, [\href{http://xxx.lanl.gov/abs/hep-th/0107138}{{\tt
  hep-th/0107138}}].

\bibitem{Aldazabal:2000dg}
G.~Aldazabal, S.~Franco, L.~E. Ib{\'a}{\~n}ez, R.~Rabad{\'a}n, and A.~M.
  Uranga, {\it D = 4 chiral string compactifications from intersecting branes},
   {\em J. Math. Phys.} {\bf 42} (2001) 3103--3126,
  [\href{http://xxx.lanl.gov/abs/hep-th/0011073}{{\tt hep-th/0011073}}].

\bibitem{Blumenhagen:2000wh}
R.~Blumenhagen, L.~G{\"o}rlich, B.~K{\"o}rs, and D.~L{\"u}st, {\it
  Noncommutative compactifications of type {I} strings on tori with magnetic
  background flux},  {\em JHEP} {\bf 10} (2000) 006,
  [\href{http://xxx.lanl.gov/abs/hep-th/0007024}{{\tt hep-th/0007024}}].

\bibitem{Angelantonj:2000hi}
C.~Angelantonj, I.~Antoniadis, E.~Dudas, and A.~Sagnotti, {\it Type-{I} strings
  on magnetised orbifolds and brane transmutation},  {\em Phys. Lett.} {\bf
  B489} (2000) 223--232, [\href{http://xxx.lanl.gov/abs/hep-th/0007090}{{\tt
  hep-th/0007090}}].

\bibitem{Chen:2005ab}
C.~M. Chen, G.~V. Kraniotis, V.~E. Mayes, D.~V. Nanopoulos, and J.~W. Walker,
  {\it A supersymmetric flipped {SU(5)} intersecting brane world},  {\em Phys.
  Lett.} {\bf B611} (2005) 156--166,
  [\href{http://xxx.lanl.gov/abs/hep-th/0501182}{{\tt hep-th/0501182}}].

\bibitem{Chen:2005cf}
C.-M. Chen, V.~E. Mayes, and D.~V. Nanopoulos, {\it Flipped {SU(5)} from
  {D}-branes with type {IIB} fluxes},  {\em Phys. Lett.} {\bf B633} (2006)
  618--626, [\href{http://xxx.lanl.gov/abs/hep-th/0511135}{{\tt
  hep-th/0511135}}].

\bibitem{Dijkstra:2004cc}
T.~P.~T. Dijkstra, L.~R. Huiszoon, and A.~N. Schellekens, {\it Supersymmetric
  standard model spectra from {RCFT} orientifolds},  {\em Nucl. Phys.} {\bf
  B710} (2005) 3--57, [\href{http://xxx.lanl.gov/abs/hep-th/0411129}{{\tt
  hep-th/0411129}}].

\bibitem{Anastasopoulos:2006da}
P.~Anastasopoulos, T.~P.~T. Dijkstra, E.~Kiritsis, and A.~N. Schellekens, {\it
  Orientifolds, hypercharge embeddings and the standard model},
  \href{http://xxx.lanl.gov/abs/hep-th/0605226}{{\tt hep-th/0605226}}.

\bibitem{Tatar:2006dc}
R.~Tatar and T.~Watari, {\it Proton decay, {Y}ukawa couplings and underlying
  gauge symmetry in string theory},
  \href{http://xxx.lanl.gov/abs/hep-th/0602238}{{\tt hep-th/0602238}}.

\bibitem{Klebanov:2003my}
I.~R. Klebanov and E.~Witten, {\it Proton decay in intersecting {D}-brane
  models},  {\em Nucl. Phys.} {\bf B664} (2003) 3--20,
  [\href{http://xxx.lanl.gov/abs/hep-th/0304079}{{\tt hep-th/0304079}}].

\bibitem{Gimon:1996rq}
E.~G. Gimon and J.~Polchinski, {\it Consistency conditions for orientifolds and
  {D}-manifolds},  {\em Phys. Rev.} {\bf D54} (1996) 1667--1676,
  [\href{http://xxx.lanl.gov/abs/hep-th/9601038}{{\tt hep-th/9601038}}].

\bibitem{Axenides:2003hs}
M.~Axenides, E.~Floratos, and C.~Kokorelis, {\it {SU(5)} unified theories from
  intersecting branes},  {\em JHEP} {\bf 10} (2003) 006,
  [\href{http://xxx.lanl.gov/abs/hep-th/0307255}{{\tt hep-th/0307255}}].

\bibitem{Berkooz:1996km}
M.~Berkooz, M.~R. Douglas, and R.~G. Leigh, {\it Branes intersecting at
  angles},  {\em Nucl. Phys.} {\bf B480} (1996) 265--278,
  [\href{http://xxx.lanl.gov/abs/hep-th/9606139}{{\tt hep-th/9606139}}].

\bibitem{Cvetic:2004nk}
M.~Cveti{\v{c}}, P.~Langacker, T.-J. Li, and T.~Liu, {\it D6-brane splitting on
  type {IIA} orientifolds},  {\em Nucl. Phys.} {\bf B709} (2005) 241--266,
  [\href{http://xxx.lanl.gov/abs/hep-th/0407178}{{\tt hep-th/0407178}}].

\bibitem{Cvetic:2003ch}
M.~Cveti{\v{c}} and I.~Papadimitriou, {\it Conformal field theory couplings for
  intersecting {D}-branes on orientifolds},  {\em Phys. Rev.} {\bf D68} (2003)
  046001, [\href{http://xxx.lanl.gov/abs/hep-th/0303083}{{\tt
  hep-th/0303083}}].

\bibitem{Lust:2004cx}
D.~L{\"u}st, P.~Mayr, R.~Richter, and S.~Stieberger, {\it Scattering of gauge,
  matter, and moduli fields from intersecting branes},  {\em Nucl. Phys.} {\bf
  B696} (2004) 205--250, [\href{http://xxx.lanl.gov/abs/hep-th/0404134}{{\tt
  hep-th/0404134}}].

\bibitem{Polchinski:1998rr}
J.~Polchinski, {\it String theory. vol. 2: Superstring theory and beyond}, .
  Cambridge, UK: Univ. Pr. (1998) 531 p.

\bibitem{Friedmann:2002ty}
T.~Friedmann and E.~Witten, {\it Unification scale, proton decay, and manifolds
  of {G(2)} holonomy},  {\em Adv. Theor. Math. Phys.} {\bf 7} (2003) 577--617,
  [\href{http://xxx.lanl.gov/abs/hep-th/0211269}{{\tt hep-th/0211269}}].

\bibitem{Lust:2003ky}
D.~L{\"u}st and S.~Stieberger, {\it Gauge threshold corrections in intersecting
  brane world models},  \href{http://xxx.lanl.gov/abs/hep-th/0302221}{{\tt
  hep-th/0302221}}.

\bibitem{Eidelman:2004wy}
{\bf Particle Data Group} Collaboration, S.~Eidelman {\em et~al.}, {\it Review
  of particle physics},  {\em Phys. Lett.} {\bf B592} (2004) 1.

\bibitem{Jung:1999jq}
C.~K. Jung, {\it Feasibility of a next generation underground water {C}herenkov
  detector: Uno},  \href{http://xxx.lanl.gov/abs/hep-ex/0005046}{{\tt
  hep-ex/0005046}}.

\bibitem{Dixon:1986qv}
L.~J. Dixon, D.~Friedan, E.~J. Martinec, and S.~H. Shenker, {\it The conformal
  field theory of orbifolds},  {\em Nucl. Phys.} {\bf B282} (1987) 13--73.

\bibitem{Arfaei:1996rg}
H.~Arfaei and M.~M. Sheikh~Jabbari, {\it Different {D}-brane interactions},
  {\em Phys. Lett.} {\bf B394} (1997) 288--296,
  [\href{http://xxx.lanl.gov/abs/hep-th/9608167}{{\tt hep-th/9608167}}].

\bibitem{Abel:2003vv}
S.~A. Abel and A.~W. Owen, {\it Interactions in intersecting brane models},
  {\em Nucl. Phys.} {\bf B663} (2003) 197--214,
  [\href{http://xxx.lanl.gov/abs/hep-th/0303124}{{\tt hep-th/0303124}}].

\bibitem{Bertolini:2005qh}
M.~Bertolini, M.~Bill{\`o}, A.~Lerda, J.~F. Morales, and R.~Russo, {\it Brane
  world effective actions for {D}-branes with fluxes},  {\em Nucl. Phys.} {\bf
  B743} (2006) 1--40, [\href{http://xxx.lanl.gov/abs/hep-th/0512067}{{\tt
  hep-th/0512067}}].

\end{thebibliography}\endgroup

\end{document}